\documentclass[pdflatex,sn-mathphys,smallcondensed]{sn-jnl}
\jyear{2021}
\usepackage{graphicx}
\usepackage{subfigure}
\theoremstyle{thmstyleone}

\usepackage{tabularx}

\newcommand{\tabincell}[2]{\begin{tabular}{@{}#1@{}}#2\end{tabular}}

\theoremstyle{thmstyletwo}

\theoremstyle{thmstylethree}

\begin{document}

\title[Long-Tail Session-based Recommendation from Calibration]{Long-Tail Session-based Recommendation from Calibration}

\author[1]{\fnm{Jiayi} \sur{Chen}}\email{jychen@ica.stc.sh.cn}
\author*[1]{\fnm{Wen} \sur{Wu}}\email{wwu@cc.ecnu.edu.cn}
\author[1]{\fnm{Liye} \sur{Shi}}\email{lyshi@ica.stc.sh.cn}
\author[2]{\fnm{Wei} \sur{Zheng}}\email{wzheng@admin.ecnu.edu.cn}
\author[1]{\fnm{Liang} \sur{He}}\email{lhe@cs.ecnu.edu.cn}

\affil[1]{\orgdiv{School of Computer Science and Technology}, \orgname{East China Normal University}, \orgaddress{\postcode{200062}, \state{Shanghai}, \country{China}}}

\affil[2]{\orgdiv{Party Working Committee}, \orgname{East China Normal University}, \orgaddress{\postcode{200062}, \state{Shanghai}, \country{China}}}

\abstract{Accurate predictions in session-based recommendations have progressed, but a few studies have focused on skewed recommendation lists caused by popularity bias. Existing models for mitigating popularity bias have attempted to reduce the overconcentration of popular items by amplifying scores of less popular items. However, they normally ignore the users' different preferences toward long-tail items. Thus, we incorporate calibration, where calibrated recommendations reflect the users' interests in recommendation lists with appropriate proportions, to mitigate the popularity bias from the user's perspective. Specifically, we propose a calibration module to predict the ratio of tail items in the recommendation list from the session representation, and align it to the ongoing session. Additionally, we utilize a two-stage curriculum training strategy to improve prediction in the calibration module. Experiments on benchmark datasets show that our model can both achieve the competitive accuracy of recommendation and provide more tail items.}


\maketitle

\section{Introduction}\label{sec1}
Recommender systems aim to help users find their interests among large-scale items. Collaborative filtering models are proposed to predict the ratings of items that users have not interacted with \cite{he2017ncf, ngcf,na21tmcf,xiao19varmf, zhang21itercf, zheng19distributedcf, huang21progress}. However, these methods take (user, item, rating) triplets as input, which does not consider the sequential features in the user's historical behaviors. In a more practical and difficult scenario, the user's information is not available if the user browses the website anonymously. Therefore, session-based recommendation (SBR) was proposed to predict the next action within a short session \cite{hidasi2015session}. Due to the practical value of SBR, many approaches focus on how to accurately predict the next action in the session. With the development of deep learning-based technologies, SBR has achieved satisfactory prediction accuracy. In general, methods that rely on Recurrent Neural Networks (RNNs) \cite{hidasi2015session,zhang21recurandconv,li2017neural, zhang21mbpi,zhang20seqgen} and Graph Neural Networks (GNNs) \cite{wu2018session,gcegnn,niser,PanCCCR20stargraph,chen20handling, qiu20crosssession, tajuddeen22gnnnonseq} contribute to the progress in SBR.

Although accurate prediction in SBR has achieved success, a few studies have attempted to mitigate the popularity bias in recommendations. Popularity bias is the imbalanced occurrence of items, where only a small proportion of items (head items) occupy the majority of occurrence according to the Pareto principle \cite{pareto}, and other less popular items are often called ``tail items''. It has been proven that popularity bias can affect recommender algorithms, resulting in a skewed recommendation list \cite{mobasher17popularitybiasltr, steck11popacc}. Such a centralized recommendation list can obtain high accuracy, but always offers a small proportion of items, which may make users feel bored \cite{tailnet}. Moreover, recommending long-tail items benefits both users and businesses \cite{08longtail, tailnet, steck18calibrated, Yin12Challenging, goel10anatomy}. Previous studies have attempted to mitigate the popularity bias in collaborative filtering models by sacrificing the prediction accuracy \cite{JohnsonN17tripartite}, while such studies on SBR are scarce. Recently, TailNet utilized a preference mechanism for long-tail SBR, which highlighted tail items in the session to enhance the score of long-tail items \cite{tailnet}. NISER amplified the tail items with the normalized representations \cite{niser}. These methods were proven effective in mitigating popularity bias in SBR.

\begin{figure}
    \centering
    \includegraphics[width=10cm]{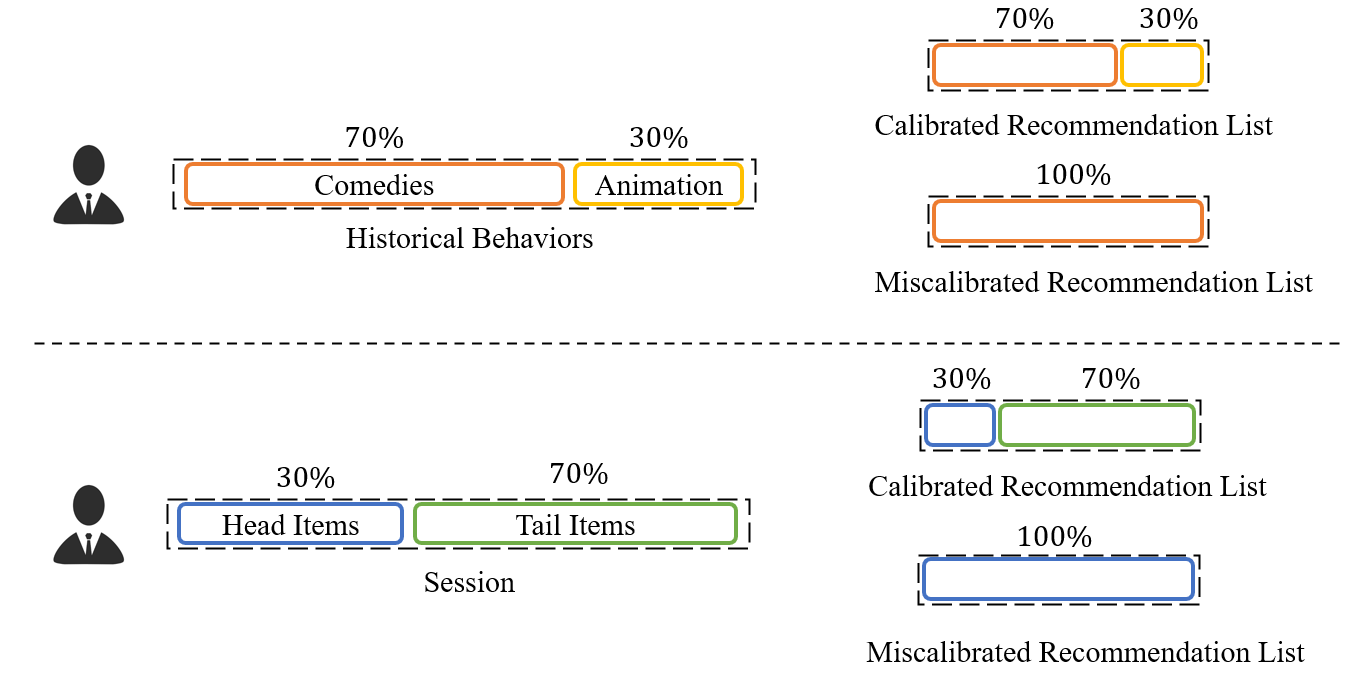}
    \caption{An illustration of calibrated recommendation}
    \label{fig:illustration}
\end{figure}

Although long-tail SBR models such as TailNet and NISER can provide more tail items, they do not consider how many tail items should be included in a recommendation list. As proposed in \cite{abdollahpouriMB21usercentered}, these methods take the item-centered perspective but ignore users' different preferences on tail items. Thus, we mitigate the popularity bias in SBR from the user perspective. We consider the long-tail SBR as a miscalibration problem, which is different from previous studies. 
A calibrated recommender system can reflect the taste (e.g., the distribution of movie genres) of a user from the historical behaviors to the recommendation list \cite{steck18calibrated, seymenAM21constrained,AbdollahpouriMB20popcalib}. As illustrated in Fig. \ref{fig:illustration}, if a user's historical records contain $70\%$ comedies and $30\%$  animations, the calibrated recommendation list should follow this distribution. It is similar to the long-tail recommendation where the proportion of tail items in the recommendation list corresponds to the historical behaviors \cite{abdollahpouriMB21usercentered}. From the perspective of calibration, the skewed recommendation list can be explained as a miscalibration problem. For the user who has viewed tail items in the past, the recommender still provides a recommendation list whose ratio of tail items is far less than the session, as shown in Fig. \ref{fig:illustration}. Recommendation lists from the user's perspective are more convincing compared to other long-tail SBR models whose ratio of tail items is not explainable nor controllable.

Another reason that causes a skewed recommendation list is the training paradigm of SBR, which was not investigated by previous studies. In the existing SBR training paradigm, sessions are treated equally and all historical behaviors are performed as targets of prediction. Such a paradigm resulted in skewed recommendation lists because targets were subjected to the long-tail distribution. 
Previous studies such as TailNet and NISER amplified scores of tail items by modifying session encoders, but they followed such a paradigm and did not investigate the inherent drawback of the SBR training paradigm. From the calibration perspective, the ratio of tail items in the ongoing session is also an objective, which still contains an imbalance problem. Because of popularity bias, only a small proportion of sessions have a high ratio of tail items, while the majority of sessions only have a few tail items. If we follow this paradigm, the imbalanced preference for tail items will further cause negative impacts, which is similar to the imbalance of item popularity. Because sessions that have a few tail items are the majority, the model designed for calibration treats the low ratio as the goal for all sessions, leading to failure in calibration. An intuitive solution is to organize the training data, which has proven effective in machine learning \cite{wang19imbalanceclassification, rasheed20rlcl} and has rarely been investigated in SBR.

Inspired by this, we propose a \textbf{C}alibrated \textbf{S}ession-\textbf{B}ased \textbf{R}ecommendation model, namely CSBR, to explore how to incorporate calibration in long-tail SBR and provide a user-centered recommendation list. First, we design a calibration module for existing SBR models. The calibration module contains two parts. One is the distribution prediction, which learns a projection from the session representation to the distribution of the recommendation list. The other is distribution alignment, which aims to align the distribution of the recommendation list toward that of the ongoing session. Then, we propose a curriculum training strategy to organize the training data and ensure the effectiveness of our calibration module. Sessions are classified into two categories by the threshold of the ratio of tail items of the session, and they are processed separately. During the evaluation stage, the recommendation list is also generated according to such a strategy. In addition, we propose an extension strategy based on threshold-based data splitting to further organize the samples.

The contributions of this paper are listed as follows:

\begin{itemize}
    \item We propose a novel CSBR model to handle the long-tail session-based recommendation from a user-centered view by incorporating calibration.
    \item We design a calibration module to align the ratio of tail items in recommendation lists and ongoing sessions. We also apply a curriculum training strategy with data splitting to organize the data and mitigate the effect of popularity bias.
    \item Experiments on benchmark datasets show that our model can both achieve competitive accuracy and recommend more tail items. In addition, the recommendation list is more calibrated compared to baseline models.
\end{itemize}
The rest of this paper is organized as follows. We review the existing literature in Sec. \ref{sec:rel}. Then we introduce our model in Sec. \ref{sec:method} and the experimental details in Sec. \ref{sec:exp}. We show the experimental results and analysis in Sec. \ref{sec:result}. Finally, we discuss our work in Sec. \ref{sec:discussion} and conclude our paper and indicate some future work in Sec. \ref{sec:conclusion}.

\section{Related Work}\label{sec:rel}
In this section, we provide a literature review of our work. We first make an introduction about session-based recommendation, and then review the recent advances in the long-tail recommendation.

\subsection{Session-based Recommendation}
Session-based recommendation has drawn attention in recent years. With the development of deep learning, neural networks have been utilized for session-based recommendation. According to the different architectures used for modeling sequences, deep learning-based methods can be divided into RNN-based models and GNN-based models. RNN-based models apply recurrent neural network and their variants, such as Gated Recurrent Unit (GRU), to model sequences. For example, Hidasi et al. proposed GRU4REC which first applied GRU in session-based recommendation, along with the ranking-based loss functions and parallel training strategy \cite{hidasi2015session}. They also proposed advanced loss functions to improve the performances of recommendation \cite{hidasi2018recurrent}. Li et al. used a hybrid global-local encoder with attention mechanisms to capture the main purpose in the session \cite{li2017neural}. Zhang et al. combined GRUs and convolution operations to capture general interest and dynamic interest \cite{zhang21recurandconv}. Additionally, RNNs have been combined with nonsequential structures to learn both sequential information and general interests \cite{zhang20seqgen}. Although RNN-based models have progressed compared to traditional models, the inherent limitation is that they can only model sequences by consecutive items and cannot learn complex item transitions in the session. Therefore, with the development of graph-based models, GNN-based models have become the trend in session-based recommendation \cite{wu2018session,ahed2020dynamicattention,PanCCCR20stargraph,cikm21ssgraphcotrain, gcegnn, tajuddeen22gnnnonseq, qiu20crosssession}. For example, Wu et al. proposed SR-GNN which treats sessions as graphs and learns item transitions and embeddings by GNNs \cite{wu2018session}. Abugabah et al. designed a dynamic attention mechanism to model the relation between the user’s historical interests and the current session \cite{ahed2020dynamicattention}. Wang et al. built a session-level graph and global-level graph to incorporate information from other sessions to enhance the recommendation performances \cite{gcegnn}. Qiu et al. also attempted to learn cross-session information by linking sessions with graphs and readout functions for enhancing session embeddings \cite{qiu20crosssession}. In addition to RNNs and GNNs, other architectures have also been applied for session-based recommendation, such as Convolutional Neural Networks \cite{tuan20173d} and Multilayer Perceptrons \cite{liu2018stamp, cao20position}.

\subsection{Long-Tail Recommendation}
The session-based recommendation models mentioned above aimed to achieve higher prediction accuracy. However, due to the popularity bias in the data, these methods tended to provide skewed recommendation lists with high popularity items. Therefore, previous studies have attempted to analyze the effect of recommending long-tail items and proved that it is beneficial to both users and businesses. For users, obtaining tail item recommendations can widen their interests and avoid the risk of the filter bubble \cite{08longtail, tailnet, steck18calibrated}. In addition, for businesses, recommending tail items can generate more profit because long-tail items earn a larger margin profit \cite{Yin12Challenging}. Providing long-tail items can also promote head item sales by ``one-stop shopping convenience''\cite{goel10anatomy}. Therefore, models were designed to mitigate the popularity bias in recommendations \cite{JohnsonN17tripartite, zhang21twotale, yan21dast, sreepada20fewshot, li21coldlongtail, ferraro19musiclongtail}. For example, Zhang et al. proposed a dual transfer framework with a meta learner that transfers knowledge from head items to tail items \cite{zhang21twotale}. Yan et al. handled long-tail recommendation via feature distribution alignment between exposed and unexposed items, and incorporated a self-training framework with pseudo labels \cite{yan21dast}. Sreepada et al. utilized few shot learning technique to mitigate the effect of popularity bias \cite{sreepada20fewshot}. Li et al. proposed a unified framework to handle both cold-start and long-tail problems \cite{li21coldlongtail}. However, previous studies mainly focused on the collaborative filtering scenario, where users' information was available and sequential behaviors were not considered. In session-based recommendation, studies on mitigating the long-tail distribution are scarce. Gupta et al. proposed normalization functions for graph-based session encoder output to diminish the total popularity of the recommendation list \cite{niser}. Liu et al. utilized a preference mechanism on RNN-based models to recommend more long-tail items by adding an extra feature vector and a preference mechanism to emphasize tail items in the session \cite{tailnet}.

In general, existing models for long-tail SBR amplified tail item scores by modifying the session encoder which generates session representations. However, the training objective and paradigm were not investigated. Users' preferences for tail items were not considered, leading to a miscalibrated ratio of long-tail items in recommendation lists. These studies also did not investigate the inherent drawback of the SBR training paradigm, where the imbalance problem caused by popularity bias was retained. Therefore, we are interested in handling long-tail SBR from the user perspective and designing a training paradigm to handle popularity bias.

\section{The CSBR Model}\label{sec:method}
In this section, we introduce our CSBR model in detail. We first provide a preliminary description of the definitions of SBR and concepts used in our CSBR model. Then we introduce our proposed calibration module and curriculum training strategy of our CSBR model. 
\subsection{Preliminaries}
We first formulate the problem of session-based recommendation. In session-based recommendation, each anonymous session, which does not contain a user ID, can be represented as an ordered list $s=\{x_{1},...,x_{n}\}$, where $x_{i} \in I$ ($I$ is the item set) is the $i-th$ clicked item out of a total number of $\|I\|$ items. All sessions construct the session set $S$, and we use $S_{tr}$ to denote the training set and $S_{te}$ as the testing set. The goal of session-based recommendation is to predict the next item $x_{n+1}$ given the current session $s=\{x_{1},...,x_{n}\}$. Following the top-N recommendation protocol, we use $RL_{s}$ to denote the recommendation list generated by session $s$. Table \ref{tab:notation} summarizes the notations. 

\begin{table}[t]
\caption{Notations used in this paper}
    \centering
    \begin{tabular}{c|l}
    \toprule
        Symbol & Description \\
    \midrule
        $I$ & The entire item set \\
        $I_{Tail}$ & All tail items \\
        $s$ & The ongoing session\\
        $x_i$ & The $i-th$ item in $s$ \\
        $x_{n+1}$ & The target item need to be predicted of session $s$ \\
        $f$ & The session encoder \\
        $h_s$ & The hidden representation of session $s$ generated by $f$ \\
        $RL_s$ & The recommendation list of session $s$ \\
        $pop(x_i)$ & The one-hot vector indicates  \\
        $q_s$ & \tabincell{l}{A 2-dimension vector which represents the popularity \\distribution of the session $s$}\\
        $p_s$ & \tabincell{l}{A 2-dimension vector which represents the popularity \\ distribution of $RL_s$} \\
        $\hat{p}_s$ & The estimated distribution used for predicting $p_s$ \\
        $\hat{q}_s$ & The estimated distribution used for the alignment to $q_s$\\
        $\lambda$ & \tabincell{l}{The weight which controls the importance of calibration \\ during model training} \\
        $S_{tr}$ & The training set\\
        $S_{te}$ & The testing set\\
        $\theta$ & The threshold of training and testing set splitting\\
        $S_{tr}(Tail)$ & \tabincell{l}{Sessions in the training set whose ratio of tail \\ items is greater than $\theta$}\\
        $S_{tr}(Head)$ & \tabincell{l}{Sessions in the training set whose ratio of tail \\ items is equal or less than $\theta$}\\
        $S_{te}(Tail)$ & \tabincell{l}{Sessions in the testing set whose ratio of tail items \\ is greater than $\theta$}\\
        $S_{te}(Head)$ & \tabincell{l}{Sessions in the testing set whose ratio of tail items \\ is equal or less than $\theta$}\\
        $K$ & Numbers of subsets for splitting $S_{tr}(Tail)$ and $S_{te}(Tail)$\\
        $index(s)$ & \tabincell{l}{The index of the subset that session $s$ belongs to.} \\
        $y_{i}$ & \tabincell{l}{The label of item $i$.}\\
        $\hat{y}_{i}$ & \tabincell{l}{The predicted probability of item $i$.}\\
    \bottomrule
    \end{tabular}
    \label{tab:notation}
\end{table}

\textbf{Head and Tail Items Distribution} Each item belongs to the head or tail subset of all items according to its popularity in $S_{tr}$. Following \cite{zhang21twotale}, we denote the top $20\%$ items with the highest popularity to head items, and others are tail items that construct the tail item set $I_{Tail}$. We use a 2-dimensional binary vector to present an item from the popularity perspective:
\begin{equation}
    pop(x)=\begin{cases}
    \{0, 1\} \quad , x \in I_{Tail} \\
    \{1, 0\} \quad , otherwise
    \end{cases}
\end{equation}
For a given session $s=\{x_{1},...,x_{n}\}$, the distribution $q_s$ of the session is calculated by:
\begin{equation}
    q_s = \frac{1}{n} \sum_{x_i \in s} pop(x_i) 
\end{equation}
where $q_s(0)$ and $q_s(1)$ are the proportions of head and tail items, respectively.

Similarly, for the recommended list $RL_{s}$ of the session $s$ generated by a session, the distribution can be calculated as follows:
\begin{equation}
    p_s = \frac{1}{N} \sum_{x \in RL_{s}} pop(x)
\end{equation}
where $N$ is the size of the recommendation list.

\subsection{Backbone Model}
The paradigm of deep learning-based models for SBR can be concluded as follows. It can be seen as a black box that takes the session as input and outputs the representation of the session and items (red box in Fig. \ref{fig:architecture}). Briefly, the procedure of the session-based recommendation is summarized as follows:
\begin{equation}
    RL_s, h_s = f(s)
\end{equation}
where $h_s$ is the session representation of session $s$ generated by the SBR model, and $f$ denotes the session encoder. By multiplying with item embeddings or multilayer perceptrons, scores of all items are computed and $RL_s$ is the recommendation list containing $N$ items with the highest scores. In our work, how to model complex item transitions to acquire higher prediction accuracy is not the main purpose. Therefore, we do not design the session encoder $f$. Instead, we apply the NARM \cite{li2017neural} and SR-GNN model \cite{wu2018session} as backbone models, which are representative models of RNNs and GNNs in SBR.

\subsection{Calibration Module}

\begin{figure}
    \centering
    \includegraphics[width=12cm]{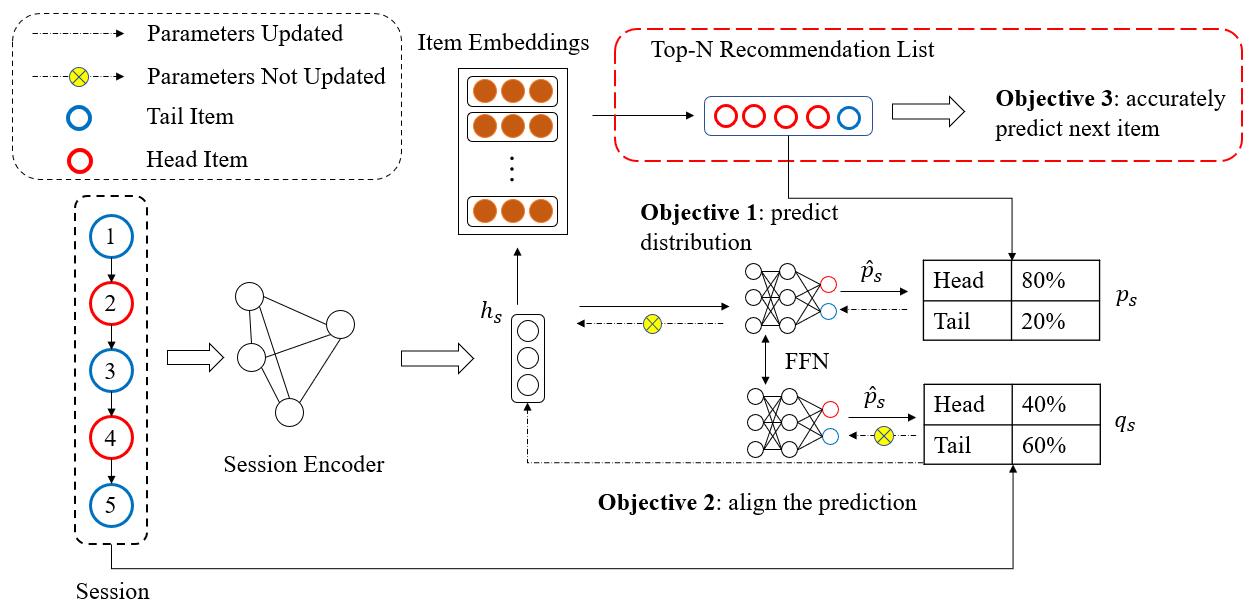}
    \caption{The architecture of the distribution prediction and align module of CSBR model. The red box represents the traditional training and recommendation paradigm of session-based recommendation. Note that only one FFN exists in our work. Here, we place two FFNs that share same parameters in this figure for better visualization. }
    \label{fig:architecture}
\end{figure}

The calibration module is shown in Fig. \ref{fig:architecture}. It contains three objectives: distribution prediction, distribution alignment and accurate next-item prediction.  

\textbf{Distribution Prediction} We take a session as the input of the backbone model $f$, and obtain the session representation $h_s$ and the recommendation list $RL_s$. The distribution $p_s$ of $RL_s$ is computed simultaneously. The goal is to learn a projection from $h_s$ to $p_s$. We use a Feed-Forward Network (FFN) to predict the distribution $p_s$ based on the hidden state $h_s$:
\begin{equation}
    \hat{p}_s = softmax(FFN(h_s))
\end{equation}
where $\hat{p}_s$ is the predicted distribution which has the same dimension as $p_s$. We choose the Kullback–Leibler (KL) divergence as the objective function to make the prediction more accurate:
\begin{align}
    L_{P}(\hat{p}_s \| p_s) &= -KL(\hat{p}_s \| p_s) \\ 
    &= - \sum_{i \in \{0,1\}}\hat{p}_s(i) \log \frac{p_s(i)}{\hat{p}_s(i)}
\end{align}
where $i \in \{0,1\}$ is the index that represents the head or tail items, respectively. In this stage, our goal is to train the FFN to predict the distribution by the session representation. Therefore, only the FFN parameters are updated while the parameters of the backbone model are fixed.

\textbf{Distribution Alignment} From the calibration perspective, the ideal distribution of recommendation list $p_s$ is close to $q_s$ to reflect the user's preference toward tail items. Therefore, we align the predicted distribution $\hat{p}_s$ generated by the same FFN toward $q_s$ with a similar loss function:
\begin{align}
    L_{Q}(\hat{p}_s \| q_s) &= -KL(\hat{p}_s \| q_s) \\ 
    &= - \sum_{i \in \{0,1\}}\hat{p}_s(i) \log \frac{q_s(i)}{\hat{p}_s(i)}
\end{align}
where $\hat{p}_s$ is the predicted distribution by the same FFN mentioned previously. Note that in this stage, the FFN parameters are fixed in order to update the parameters of the backbone model under the calibration constraint. 

Item popularity bias may result in session popularity bias, where sessions with low $q_s(1)$ values occupy the majority. The imbalanced distribution will cause negative impacts on our calibration module, where the calibration module provides skewed outputs for all sessions. To further handle the imbalance problem in distribution, we add a weight based on the calibration loss function:
\begin{align}\label{eq:weight}
    L_{WP} &= p_s(1) \cdot L_{P}(\hat{p}_s\|p_s)\\
    L_{WQ} &= q_s(1) \cdot L_{Q}(\hat{p}_s\|q_s)
\end{align}
In this weighted loss function, we aim to reward the recommendation lists and sessions that contain a high ratio of tail items by multiplying by the ratio. Therefore, the value of the loss function is enhanced for these samples.

\textbf{Accurate Prediction} In addition to calibration loss, the goal of accurate prediction is still important. Therefore, we preserve the cross entropy loss function in the calibration module. The final loss function can be written as:
\begin{equation}\label{eq:lambda}
    L = L_{CE} + \lambda (L_{WP} + L_{WQ})
\end{equation}
where $\lambda$ is a weighting factor that controls the importance of calibration in the training procedure. $L_{CE}$ is a commonly used cross entropy loss function:
\begin{equation}
    L_{CE} = - \sum_{i=1}^{\|I\|} y_{i}\log (\hat{y}_i)
\end{equation}
where $y_{i}=1$ if item $i$ is the next item $x_{n+1}$ of session $s$; otherwise $y_{i}=0$. $\hat{y}_i$ is the predicted probability of item $i$, which is normalized by a softmax function over all items.

\subsection{Curriculum Training}\label{sec:curriculum-training}
In this section, we introduce the training procedure of our model and the generation of a calibrated recommendation list. Specifically, we propose a curriculum training strategy with a threshold-based data organization and a pre-training and fine-tuning procedure. Moreover, we introduce the extension of threshold-based data organization.

\begin{figure}
    \centering
    \includegraphics[width=10cm]{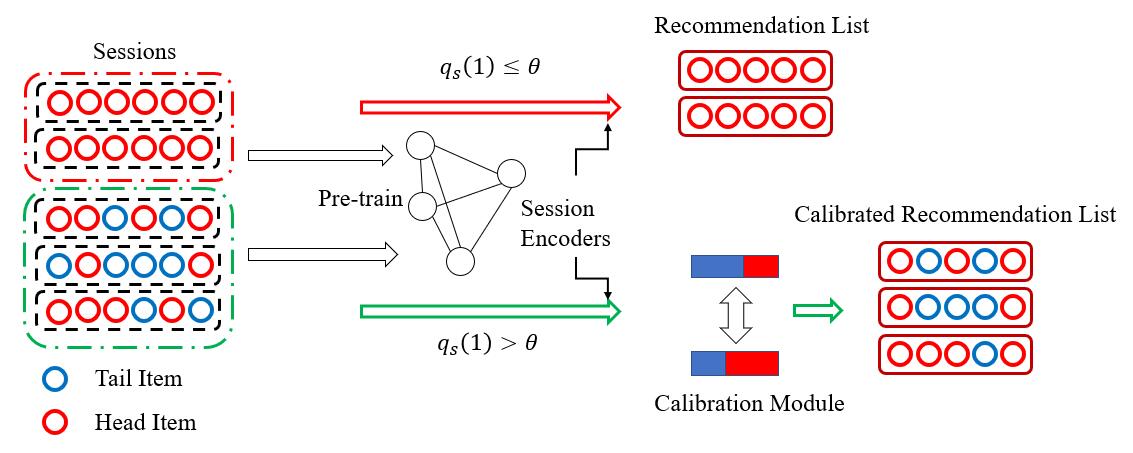}
    \caption{The curriculum training strategy. Sessions are processed according to the ratio of tail items.}
    \label{fig:curriculum-training}
\end{figure}

\textbf{Threshold-based Data Organization} In our model, the ratio of tail items in the ongoing session performs as the goal of the calibration module. Due to popularity bias, the ratio is not subject to a balanced distribution, where sessions with low $q_s(1)$ occupy the majority. Therefore, it is necessary to organize the training sessions to avoid imbalance. As shown in Fig. \ref{fig:curriculum-training}, we split the training set into two categories according to the sessions' $q_s(1)$ value:
\begin{equation}
    s \in \begin{cases}
    S_{tr}(Tail) \quad , q_s(1) \textgreater \theta \\
    S_{tr}(Head) \quad , otherwise
    \end{cases}
\end{equation}
which means that if the ratio of tail items in session $s$ is greater than $\theta$, it belongs to the tail subset $S_{tr}(Tail)$, and the head subset $S_{tr}(Head)$ otherwise. When $\theta=0$, sessions that do not have tail items are moved to $S_{tr}(Head)$. Only sessions in $S_{tr}(Tail)$ are further processed by our calibration module.

\textbf{Training and evaluation} We first assign two backbone models for the two subsets $S_{tr}(Tail)$ and $S_{tr}(Head)$. We train the parameters of the two models on the entire training set $S_{tr}$ to learn item transition information following the traditional sequence-to-item training paradigm of SBR \cite{li2017neural, wu2018session, niser}. Then we optimize the model by our calibration algorithm on $S_{tr}(Tail)$. Note that for $S_{tr}(Head)$, there is no further execution so that the outputs of sessions in $S_{tr}(Head)$ are the same as the backbone model. 

The testing set is also split into $S_{te}(Tail)$ and $S_{te}(Head)$ by following the criteria above. For recommendation and evaluation, given a session $s$, the recommendation list is also based on its ratio of tail items. If $s \in S_{te}(Tail) $, the output is from our optimized model. Otherwise, the output is from the original backbone model. Finally, the top-N recommendation list is generated. 

\textbf{K-fold Extension} The above threshold-based data organization splits the dataset into two subsets according to the $q_{s}(1)$ value. In addition to splitting the whole dataset, we further investigate the data organization on the tail subset generated by the above data organization approach. We divide subsets $S_{tr}(Tail)$ and $S_{te}(Tail)$ into fine-grained groups according to the $q_{1}(1)$ value. We split sessions into $K$ different subsets according to the ratio of tail items (i.e., the value of $q_s(1)$):
\begin{equation}\label{eq:split}
    index(s) = \lceil q_s(1) \times K \rceil 
\end{equation}
where $index(s) \in \{1,2,...,K\}$ is the subset index to which session $s$ belongs, and $\lceil v \rceil$ represents the ceil function that provides the lowest integer $t$ that satisfies $t \ge v$. Note that in the K-fold extension, we consider all sessions which have at least one tail item without the threshold $\theta$. We assign one backbone model for each subset, and there are $K+1$ models including the original backbone model. For session $s$, the output comes from the corresponding model:
\begin{equation}
    h_s, RL_s = f_{index(s)}(s)
\end{equation}
The training and evaluation are similar to the above processes, where all $K$ session encoders are optimized separately and do not share parameters.

\section{Experiments}\label{sec:exp}
To demonstrate the effectiveness of our model, we conducted experiments on real-world datasets and compared our model with existing models. 
\subsection{Dataset}
We adopted two commonly-used benchmark datasets to evaluate the performances of our model. The first one is \emph{YOOCHOOSE}\footnote{http://2015.recsyschallenge.com/challenge.html} which was released in Recsys 15 and collected from a e-commerce platform. The other is \emph{Last.fm}\footnote{http://ocelma.net/MusicRecommendationDataset/Last.fm-1K.html} which is used for music artist recommendations. We further preprocessed the original datasets following \cite{li2017neural, chen20handling, gcegnn}. The statistics are listed in Table \ref{table-statistic}.

\begin{table}[t]
\centering
\caption{Statistics of Datasets}
\begin{tabular}{cccc}
    \hline
     Statistics  & YOOCHOOSE & Last.fm\\
     \hline
     Number of Clicks & 8,326,407 & 3,835,706\\
     Number of Training Sessions & 5,917,745 & 2,837,644  \\
     Number of Test Sessions  & 55,898 & 672,519 \\
     Number of Items  & 29,618 & 38,615 \\
     Average Session Length  & 5.71 & 9.19 \\
     \hline
\end{tabular}
\label{table-statistic}
\end{table}

\subsection{Comparison Models}
We selected the following methods as baselines:
\begin{itemize}
    \item \textbf{GRU4REC} \cite{hidasi2015session}, which applies RNNs to model sequential behaviors and employs ranking-based loss functions.
    \item \textbf{NARM} \cite{li2017neural} utilizes an attention mechanism with RNNs to learn user's main purpose and sequential behavior. 
    \item \textbf{SR-GNN} \cite{wu2018session} is a state-of-the-art SBR model that applies graph neural networks to learn item transitions. 
    \item \textbf{TailNet} \cite{tailnet} designs a preference mechanism for long-tail session-based recommendation.
    \item \textbf{NISER} \cite{niser} applies normalization methods to mitigate the long-tail effect.
\end{itemize}
Our CSBR model includes two types according to the selection of backbone models: 
\begin{itemize}
    \item \textbf{CSBR(N)} applies NARM as its backbone model.
    \item \textbf{CSBR(S)} applies SR-GNN as its backbone model. 
\end{itemize}

\subsection{Evaluation Metrics}\label{sec:metrics}
We evaluate models from accuracy and long-tail perspectives. Accuracy-based metrics measure the model performance by the ranking position of the user's next behavior in the recommendation list. Following previous work\cite{wu2018session, li2017neural}, we use Recall and MRR as evaluation metrics. 
\begin{itemize}
    \item \textbf{Recall@N} (Rec@N) is a widely used metric in recommendation and information retrieval areas. Recall@N computes the proportion of correct items in the top-N items of the list.
    \begin{equation}
        Recall@N = \frac{1}{\|S_{te}\|} \sum_{s \in S_{te}} 1(x_{n+1} \in RL_{s})
    \end{equation}
    where $1()$ is an indication function whose value equals 1 when the condition in brackets is satisfied and 0 otherwise. 
    \item \textbf{MRR@N} is another important metric that considers the rank of correct items. The score is computed by the reciprocal rank when the rank is within N; otherwise the score is 0. 
    \begin{equation}
        MRR@N = \frac{1}{\|S_{te}\|} \sum_{s \in S_{te}} \frac{1}{rank(x_{n+1}, RL_{s})}
    \end{equation}
\end{itemize}

The long-tail based metrics measure the performance of how many tail items can be recommended to users. Following \cite{tailnet}, we use Coverage@N, TailCoverage@N, and Tail@N as the evaluation metrics:
\begin{itemize}
    \item \textbf{Coverage@N} (Cov@N) is a widely used metric. It represents the proportion of recommended items to the entire item set.
    \begin{equation}
        Coverage@N = \frac{\|\bigcup_{s \in S_{te}} RL_{s}\|}{\|I\|} 
    \end{equation}
    \item \textbf{TailCoverage@N} (TCov@N) is similar to $coverage$ but computes only the proportion of tail items that are recommended on the testing set.
    \begin{equation}
        TailCoverage@N = \frac{\|\bigcup_{s \in S_{te}} (RL_{s} \cap I_{Tail} )\|}{\|I_{Tail}\|} 
    \end{equation}
    \item \textbf{Tail@N} is the proportion of tail items in the recommendation list for each session:
    \begin{equation}
        Tail@N = \frac{1}{\|S_{te}\|} \sum_{s \in S_{te}} \frac{\|RL_{s} \cap I_{Tail} \|}{N} 
    \end{equation}
\end{itemize}
In addition, to make intuitive comparisons, we calculate the improvement of our model against other models as follows:
\begin{equation}
     IP = \frac{Metric_{ourmethod} - Metric_{othermethod}}{Metric_{othermethod}}
\end{equation}
where $Metric$ can be any one of the metrics above.

\subsection{Experimental Setup}
We use grid search to find the optimal hyperparameters on the validation dataset which is $10\%$ of the training set. The dimension of all hidden states is set to 100, following \cite{wu2018session}. We tune the learning rate from $\{0.1,0.01,0.001\}$. We use the Adam \cite{adam} optimizer with a batch size of 128. We report the performances under the model parameters with the optimal prediction accuracy on the validation set. For the hyperparameter $\lambda$, $K$ and $\theta$, we report the performance when $\lambda=1$, $K=1$ and $\theta=0$ which is the default setting of our model. We further analyze the performance change under different hyperparameters in the following sections. For the top-N recommendation, we set $N=20$ following \cite{li2017neural, wu2018session, tailnet}.

\section{Results and Analysis}\label{sec:result}
We aim to answer the following research questions by conducting the experiments:
\begin{itemize}
    \item \textbf{RQ1}: Does our CSBR model outperform the state-of-the-art session-based recommendation models?
    \item \textbf{RQ2}: Does our CSBR model generate a more calibrated recommendation list compared to the baselines?
    \item \textbf{RQ3}: How well does the CSBR model perform when hyperparameters change?
    \item \textbf{RQ4}: How do the calibration module and curriculum training strategy of the CSBR model influence the performance?
\end{itemize}
\begin{table}[tbp]
  \centering
  \caption{Performances on the YOOCHOOSE dataset (the best results are marked in bold).}
    \begin{tabular}{c|ccccc}
    \toprule
    YOOCHOOSE & Rec@20 & MRR@20 & Cov@20 & TCov@20 & Tail@20 \\
    \midrule
    GRU4REC  & 66.81  & 27.75 & 28.51  & 11.40  & 3.86  \\
    \midrule
    NARM  & 68.85  & 28.69 & 42.29  & 28.35  & 5.76  \\
    \midrule
    SR-GNN  & 71.45  & 31.85  & 42.35  & 28.29  & 5.66  \\
    \midrule
    TailNet & 68.25  & 30.55  & 47.16  & 34.17  & 6.43  \\
    \midrule
    NISER & \textbf{71.52 } & \textbf{32.17 } & 46.56  & 33.56  & 5.95  \\
    \midrule
    CSBR(N) & 67.55 & 27.97 & 49.42 & 37.32 & 6.71 \\
    \midrule
    CSBR(S) & 71.44  & 31.92  & \textbf{50.85 } & \textbf{38.96 } & \textbf{6.73 } \\
    $Improv.$ & -$0.11\%$ & -$0.78\%$ & +$7.82\%$ & +$14.01\%$ & +$4.67\%$ \\
    \bottomrule
    \end{tabular}
  \label{tab:performance-y4}
\end{table}

\begin{table}[tbp]
  \centering
  \caption{Performances on the Last.fm dataset (the best results are marked in bold).}
    \begin{tabular}{c|ccccc}
    \toprule
    Last.fm & Rec@20 & MRR@20 & Cov@20 & TCov@20 & Tail@20 \\
    \midrule
    GRU4REC  & 19.88  & 6.55  & 27.32 & 10.99 & 1.67  \\
    \midrule
    NARM  & 21.05  & 6.95  & 51.77  & 40.34  & 3.84  \\
    \midrule
    SR-GNN  & \textbf{22.44 } & \textbf{8.88 } & 67.44  & 59.31  & 4.11  \\
    \midrule
    TailNet & 20.31  & 7.35  & 65.16  & 57.29  & 2.99  \\
    \midrule
    NISER & 22.22  & 8.74  & 76.45  & 70.56  & 4.83  \\
    \midrule
    CSBR(N) & 21.53 & 7.02  & 71.90  & 65.29 & 5.57 \\
    \midrule
    CSBR(S) & 21.65 & 8.60   & \textbf{78.60} & \textbf{73.25} & \textbf{5.74} \\
    $Improv.$ & $-3.52\%$ & $-3.15\%$ & +$2.15\%$ & +$3.81\%$ & +$18.83\%$ \\
    \bottomrule
    \end{tabular}
  \label{tab:performance-Last.fm}
\end{table}

\subsection{RQ1: Overall Performances}\label{sec:overall}
The performances of baselines and our model are listed in Table \ref{tab:performance-y4} and Table \ref{tab:performance-Last.fm}\footnote{For better visualization, we omit the symbol $\%$ when listing performances in tables. In addition, the improvements in the tables are compared between our CSBR(S) model and the best performance in the baseline models.}, where the best performances are marked in bold. In general, our CSBR model achieves competitive prediction accuracy compared to the state-of-the-art SBR models, and outperforms these baseline models in terms of long-tail-based metrics on two datasets.

We first analyze the performances from the perspective of accurate recommendation (i.e., Rec@20 and MRR@20). On the YOOCHOOSE dataset, our CSBR(S) model is slightly worse than the strongest baseline NISER model ($71.44\%$ vs. $71.52\%$ on Rec@20 and $31.92\%$ vs. $32.17\%$ on MRR@20). On the Last.fm dataset, the SR-GNN model achieves the best performance, and our CSBR(S) model is also competitive ($21.65\%$ vs. $22.44\%$ on Rec@20 and $8.60\%$ vs. $8.88\%$ on MRR@20). Although the objective of calibration is included, our CSBR model still achieves high Rec@20 and MRR@20. A possible reason is that our model follows a pre-training and fine-tuning paradigm, which learns item transitions in advance. Moreover, the calibration module of our CSBR model contains a cross entropy loss function, which ensures recommendation accuracy. 

For long-tail metrics, our CSBR framework outperforms baseline models on two datasets. On the YOOCHOOSE dataset, our CSBR(S) model achieves $50.85\%$ and $38.96\%$ of Cov@20 and TCov@20, respectively, which are higher than the best in baselines ($47.16\%$ and $34.17\%$ of TailNet). On the Last.fm dataset, the values of Cov@20 and TCov@20 are $78.6\%$ and $73.25\%$ for our CSBR(S) model, respectively. NISER obtains the best performance in the baselines, in which Cov@20 and TCov@20 are $76.45\%$ and $70.56\%$, respectively, and our model is better than NISER. In addition to the coverage over the itemset, the Tail@20 measures the proportion of tail items in the recommendation lists. On two datasets, our CSBR(S) model also outperforms the baselines (e.g., $5.74\%$ vs. $4.83\%$ of CSBR(S) and NISER on the Last.fm dataset with an improvement of $18.84\%$). The comparisons demonstrate that our model can recommend more tail items to users. Note that metrics such as Cov@20 and TCov@20 evaluate the occurrence of all items binarily. For a long-tail item, there is no difference whether it is recommended once or ten times from the coverage perspective. However, Tail@20 measures the proportion of tail items in the recommendation lists, which represents the total occurrence of long-tail items. From this perspective, the long-tail recommendation ability of our model is more convincing. 

In addition to the comparisons between our CSBR model and the state-of-the-art baselines, we also compare two variants of our model that contain different backbone models (i.e., CSBR(S) and CSBR(N)). Both of them obtain improvement compared to their corresponding backbone models. Compared with the graph-based backbone models, our CSBR(S) achieves higher performance than SR-GNN on long-tail-based metrics on two datasets. For example, on the YOOCHOOSE dataset, our CSBR(S) model obtains an improvement of $37.71\%$ on TCov@20 and $18.90\%$ on Tail@20. On the Last.fm dataset, the improvements of our model are $23.50\%$ and $39.66\%$ in terms of TCov@20 and Tail@20, respectively. Similarly, our CSBR(N) model also achieves better performance than its backbone model NARM. For example, NARM achieves $51.77\%$, $40.34\%$ and $3.84\%$ in terms of Cov@20, TCov@20 and Tail@20 on the Last.fm dataset. The performances of Cov@20, TCov@20 and Tail@20 are $71.90\%$, $65.29\%$, and $5.57\%$ for our CSBR(N), with the improvements of $38.88\%$, $61.84\%$, and $45.05\%$, respectively. The performance comparisons show that our model can mitigate the popularity bias and expose more tail items compared to the existing SBR models which do not consider popularity bias. 

We also focus on the difference between the performances on the two datasets. On the YOOCHOOSE dataset, the results of Rec@20 and MRR@20 are approximately $70\%$ and $30\%$, respectively. However, on the Last.fm dataset, the accuracy performances are approximately $21\%$ and $8\%$. For long-tail-based metrics, the performances on the Last.fm dataset are better than on the YOOCHOOSE dataset. Taking Cov@20 as an example, SR-GNN achieves $67.44\%$ on the Last.fm dataset, which is higher than $42.35\%$ on the YOOCHOOSE dataset. A possible reason is that the sequential patterns of behaviors on the YOOCHOOSE dataset are more fixed than those on the Last.fm dataset. Fixed sequential patterns make recommendations more determinate and make it easier to obtain higher accuracy. In contrast, relatively loose sequential patterns make it difficult to achieve high accuracy, but more items are covered in the recommendations.

We display and analyze the overall performances in Tables 3 and 4, where all sessions in $S_{te}$ are evaluated. However, our CSBR model works on the sessions whose ratio of tail items is greater than $\theta$ (see Sec. \ref{sec:curriculum-training}) and follows the original outputs of the backbone model for other sessions. Therefore, we compare the performances of our CSBR(S) model with the backbone SR-GNN model and the state-of-the-art NISER model on the subset $S_{te}(Tail)$ to further investigate the improvement of our model. The performance comparisons are shown in Table \ref{tab:subset} and $\theta$ equal to zero following the previous setting. 

\begin{table}[tbp]
  \centering
  \caption{The performance comparisons on the $S_{te}(Tail)$ of the testing set. The best performances are marked in bold.}
    \begin{tabular}{c|c|ccccc}
    \toprule
    Dataset & Methods & Rec@20 & MRR@20 & Cov@20 & TCov@20 & Tail@20 \\
    \midrule
    \multirow{4}[4]{*}{YOOCHOOSE} & SR-GNN & 59.67 & 23.85 & 40.10  & 26.83 & 22.28 \\
          & NISER & \textbf{60.07} & \textbf{24.50}  & 42.98 & 30.37 & 23.29 \\
\cmidrule{2-7}          & CSBR(S) & 59.33 & 23.83 & \textbf{48.22} & \textbf{37.65 }& \textbf{28.92} \\
          & Improv. & -1.23\% & -2.73\% & +12.19\% & +23.97\% & +24.17\% \\
    \midrule
    \multirow{4}[4]{*}{Last.fm} & SR-GNN & \textbf{19.91} & \textbf{7.58}  & 68.47 & 60.61 & 6.85 \\
          & NISER & 19.65 & 7.22  & 76.11 & 70.16 & 7.65 \\
\cmidrule{2-7}          & CSBR(S) & 18.64 & 7.10   & \textbf{78.45} & \textbf{73.13 }& \textbf{9.70} \\
          & Improv. & -6.37\% & -6.33\% & +3.07\% & +4.23\% & +26.79\% \\
    \bottomrule
    \end{tabular}
  \label{tab:subset}
\end{table}

For accurate prediction (i.e., Rec@20 and MRR@20), our CSBR(S) model is close to SR-GNN and NISER in general. On the YOOCHOOSE dataset, the Rec@20 are $59.33\%$, $60.07\%$ and $59.67\%$ for CSBR(S), NISER and SR-GNN, respectively. On the Last.fm dataset, the Rec@20 are $18.64\%$, $19.65\%$ and $19.91\%$ for CSBR(S), NISER and SR-GNN, respectively. The relatively looser sequential patterns of the Last.fm dataset result in a larger decrease in recommendation accuracy than the YOOCHOOSE dataset. In terms of long-tail recommendation, our CSBR model achieves more improvement compared to Tables 3 and 4. On the YOOCHOOSE dataset, the improvement of our model is $14.01\%$ in terms of TCov@20, while it is $23.97\%$ on $S_{te}(Tail)$ as shown in Table \ref{tab:subset}. In addition, the improvements of Tail@20 on all sessions and $S_{te}(Tail)$ are $4.67\%$ and $24.17\%$, respectively. On the Last.fm dataset, the improvements on $S_{te}(Tail)$ are also larger than those on all sessions. For example, the Tail@20 of our model is $5.74\%$ in Table \ref{tab:performance-Last.fm}, while it is $9.70\%$ on $S_{te}(Tail)$ in Table \ref{tab:subset}. The performances on $S_{te}(Tail)$ demonstrate the ability of our model to recommend more tail items for users. 

\subsection{RQ2: Calibrated Recommendation}
\begin{figure}[tbp]
\centering
\subfigure[Performances of $C_{KL}@20$ under different levels of $q_{s}(1)$ on the YOOCHOOSE dataset.]{
\includegraphics[width=5cm]{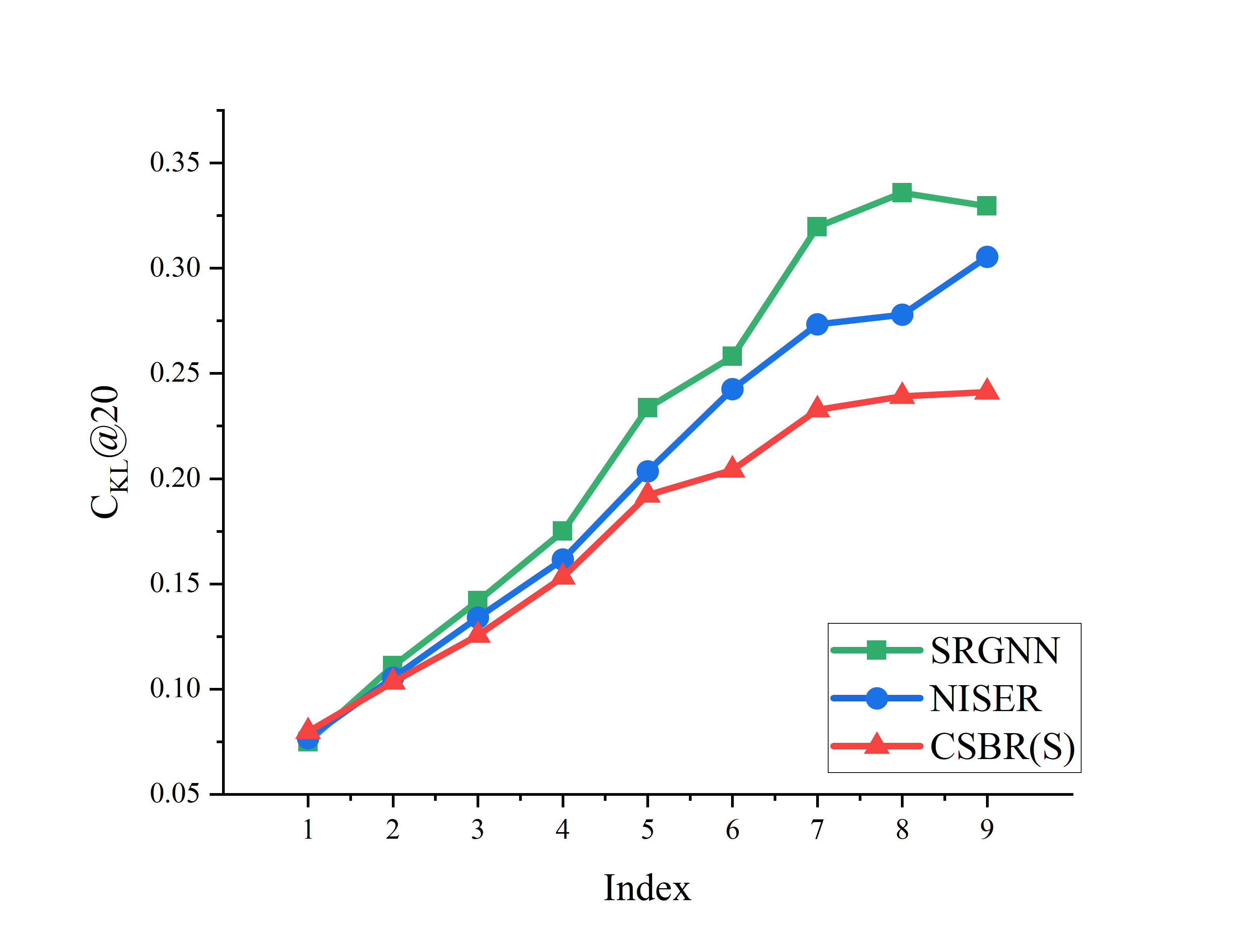}
}
\quad
\subfigure[Performances of Tail@20 under different levels of $q_{s}(1)$ on the YOOCHOOSE dataset.]{
\includegraphics[width=5cm]{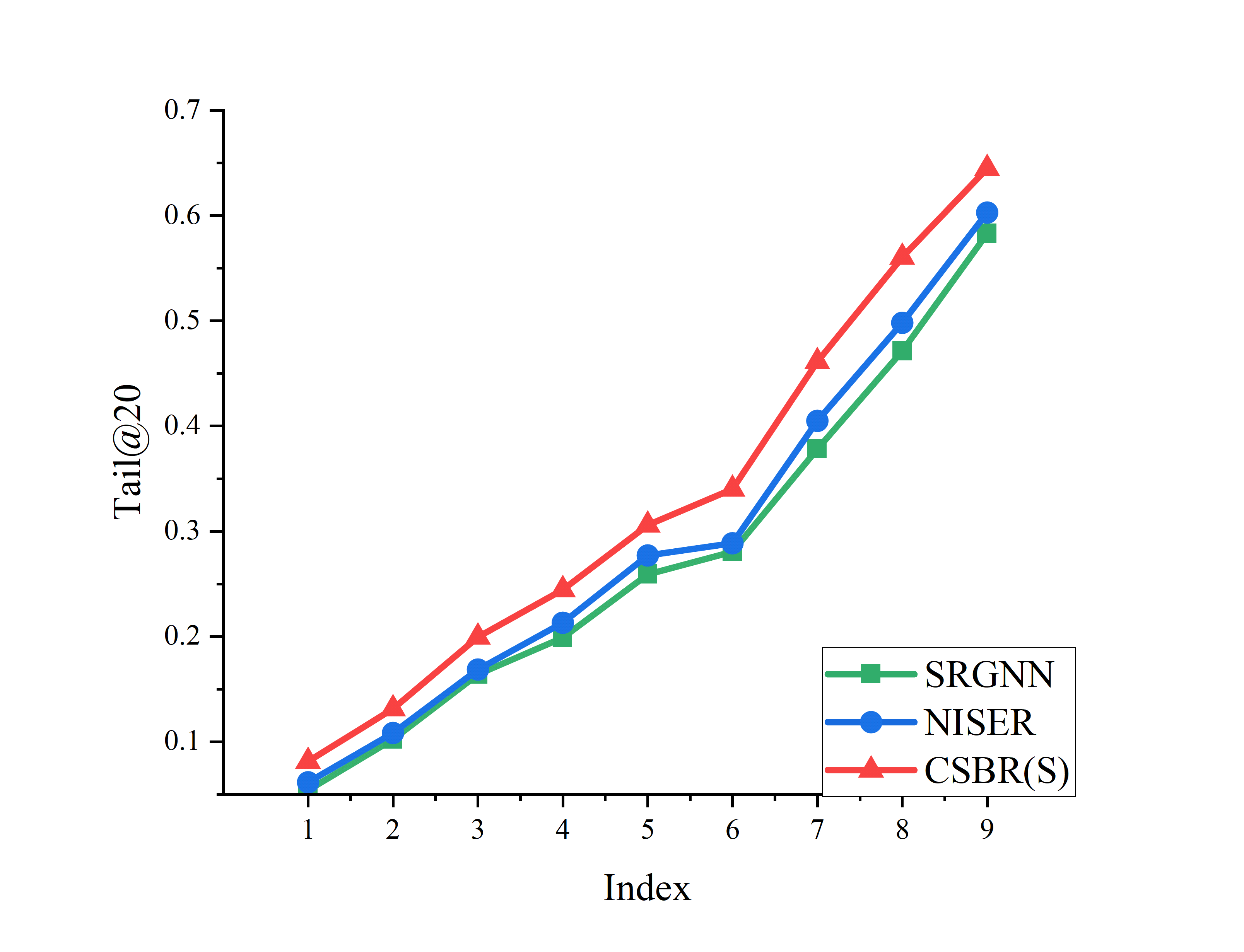}
}
\quad
\subfigure[Performances of $C_{KL}@20$ under different levels of $q_{s}(1)$ on the Last.fm dataset.]{
\includegraphics[width=5cm]{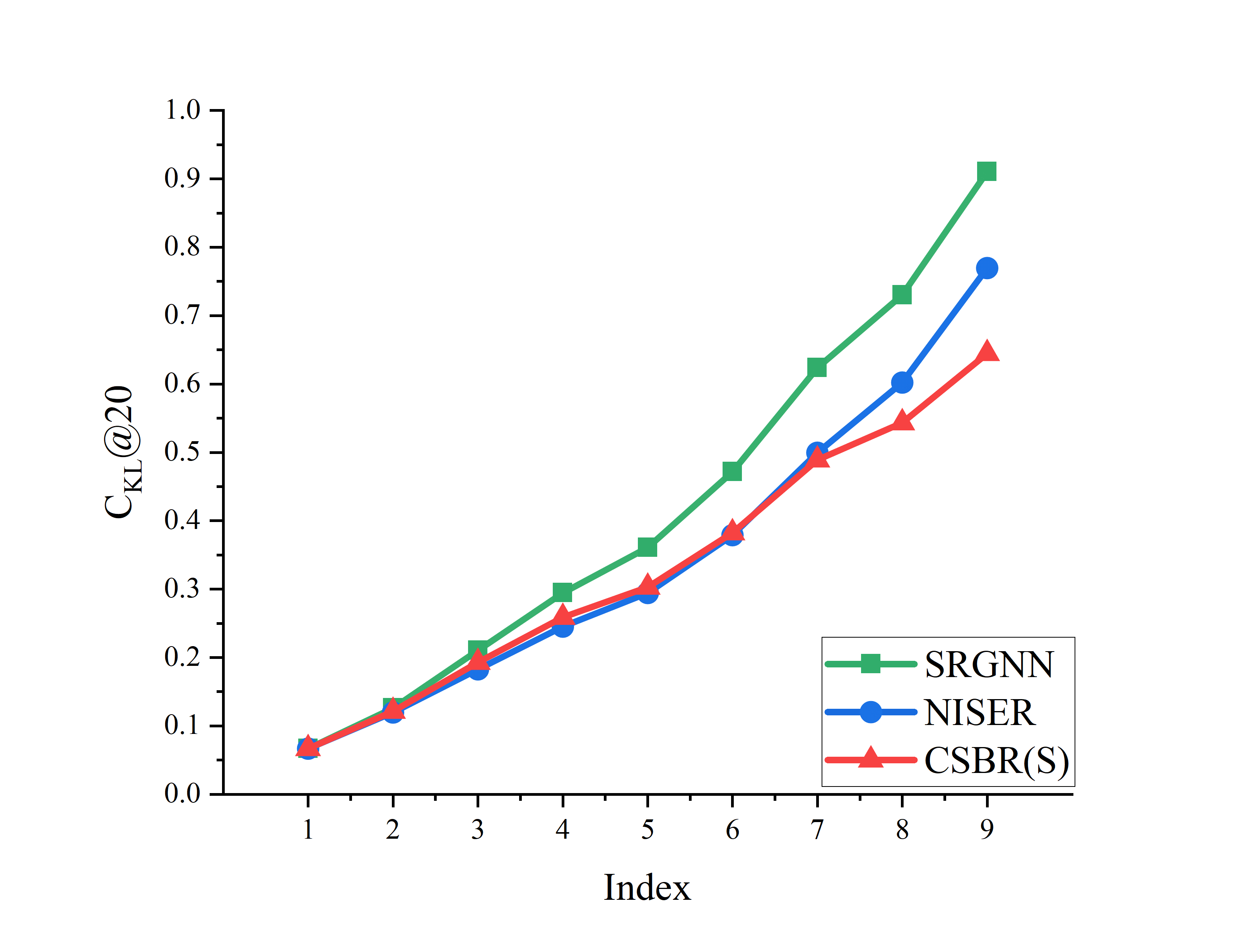}
}
\quad
\subfigure[Performances of Tail@20 under different levels of $q_{s}(1)$ on the Last.fm dataset.]{
\includegraphics[width=5cm]{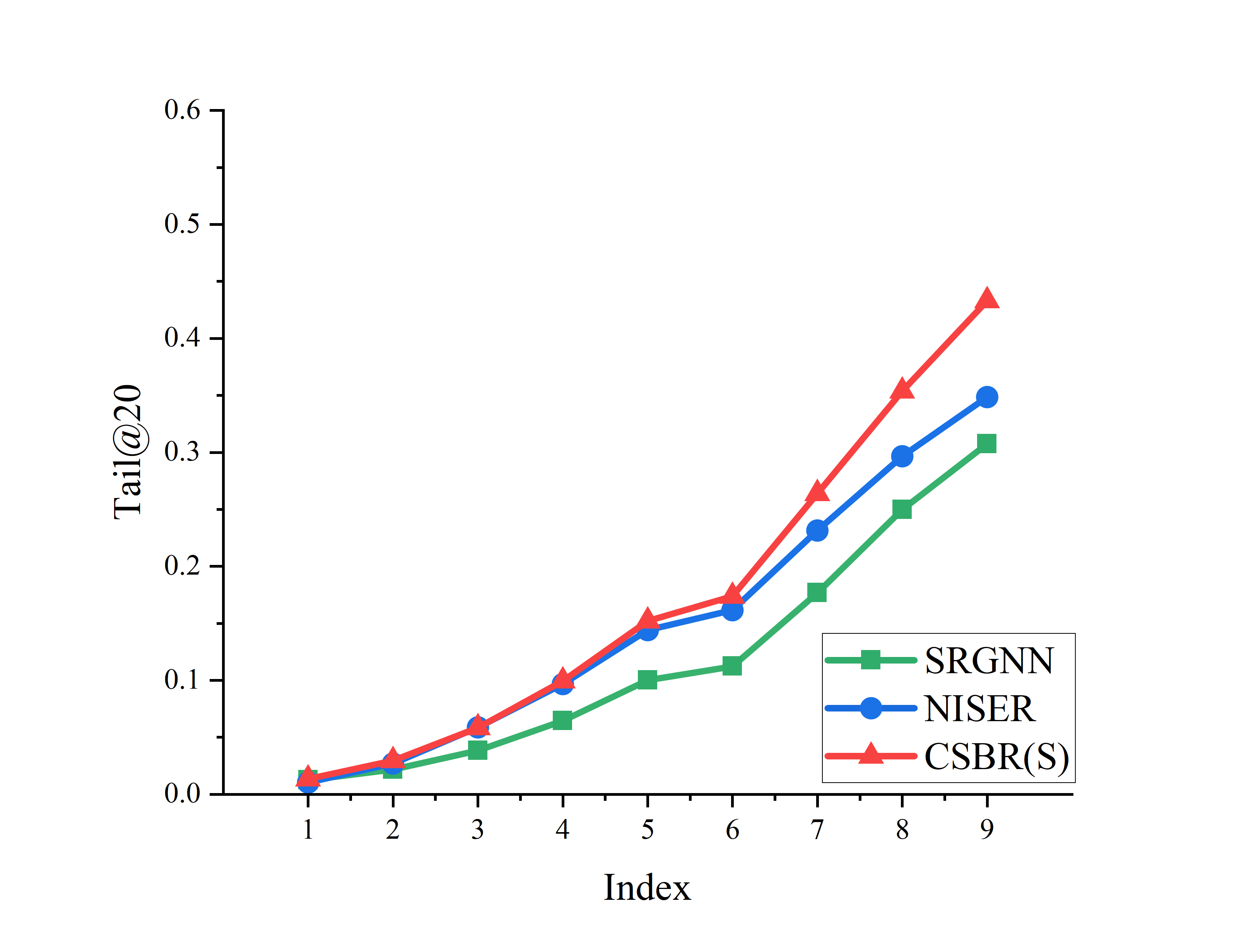}
}

\caption{Performances of $C_{KL}$ and Tail@20 under different levels of $q_{s}(1)$ value. }
\label{fig:qlevel}
\end{figure}

In this section, we answer question RQ2 about the calibration recommendation. To evaluate the calibration, we apply the $C_{KL}$ \cite{steck18calibrated} metric to compute the KL divergence between the recommendation list distribution and session distribution:
\begin{align}
    C_{KL} &= \frac{1}{\|S_{te}(Tail)\|}\sum_{s \in S_{te}(Tail)} KL(p_s \| q_s) \\
           &= \frac{1}{\|S_{te}(Tail)\|}\sum_{s \in S_{te}(Tail)} \sum_{i \in \{0, 1\}} p_s(i) \log \frac{q_{s}(i)}{p_s(i)}
\end{align}
The value of $C_{KL}$ represents the consistency of the two distributions. In our work, a lower $C_{KL}$ means that more calibrated recommendation lists are generated. For example, if $C_{KL} = 0$, the ratio of tail items in the recommendation list matches perfectly with the ongoing session. We compare our CSBR(S) model, its backbone SR-GNN model, and the state-of-the-art NISER model in terms of $C_{KL}$. Specifically, we split the tail subset $S_{te}(Tail)$ following Eq. \ref{eq:split} and set $K=10$ to evaluate the performances under different levels of $q_{s}(1)$. The comparisons are shown in Fig. \ref{fig:qlevel}. Cases whose $q_s(1)$ value equals 1 and $q_s(1)$ equals to 0 lead to the error in calculating $C_{KL}$. Therefore, we set $q_s(1)$ to 0.999 and $q_s(0)$ to 0.001 for these cases and list the performances in Table \ref{tab:qs1} separately.

Compared to the backbone SR-GNN model, our CSBR(S) model contributes to the long-tail recommendation by achieving lower $C_{KL}$ and higher Tail@20 at each level of the $q_s(1)$ value. For example, on the YOOCHOOSE dataset, the Tail@20 of our model and the SR-GNN model at the level index 3 are $0.1995$ and $0.1642$, respectively. For sessions with a high level of $q_s(1)$, our model also obtains better performances than the SR-GNN model (e.g., 5.6570 v.s. 7.1842 of $C_{KL}$ on the Last.fm dataset as listed in Table \ref{tab:qs1}). The results demonstrate the effectiveness of our core idea of incorporating calibration for the long-tail recommendation. 

Compared to the NISER model, which is one of the state-of-the-art methods for the long-tail session-based recommendation, our CSBR(S) differs from it at the high $q_s(1)$ level. For example, on the Last.fm dataset, lines representing the CSBR(S) and NISER model almost overlap until index 7. As listed in Table \ref{tab:qs1}, the $C_{KL}$ of our model is 5.6570 while it is 6.5187 for NISER. Meanwhile, Tail@20 of our model is higher than NISER (0.3468 v.s. 0.2369). The NISER model provides more tail items by normalizing item and session embeddings, leading to the improvement of $C_{KL}$ at low levels of the $q_s(1)$ value. However, its ability to deal with sessions containing a high ratio of long-tail items is weaker than our CSBR(S) model. By incorporating calibration, a higher ratio of long-tail items in recommendation lists becomes a goal of our model, resulting in the improvement in both calibration and long-tail session-based recommendation. 
\begin{table}[tbp]
  \centering
  \caption{Performances of the subset whose index is 10 in terms of $C_{KL}$ and Tail@20. The best performances are marked in bold.}
    \begin{tabular}{c|cc|cc}
    \toprule
    \multirow{2}[2]{*}{ } & \multicolumn{2}{c|}{YOOCHOOSE} & \multicolumn{2}{c}{Last.fm} \\
          & $C_{KL}$   & Tail@20 & $C_{KL}$ & Tail@20 \\
    \midrule
    SR-GNN & 4.9206 & 0.4079 & 7.1842 & 0.1787 \\
    NISER & 4.7278 & 0.4262 & 6.5187 & 0.2369 \\
    CSBR(N) & \textbf{3.3177} & \textbf{0.5724} & \textbf{5.6570} & \textbf{0.3468} \\
    \bottomrule
    \end{tabular}
  \label{tab:qs1}
\end{table}

\subsection{RQ3: Parameter Influence}

We use three hyperparameters in our CSBR framework. The parameter $\lambda$ (see Eq. \ref{eq:lambda}) controls the importance of the calibration loss function. The parameter $\theta$ is the threshold for training and test dataset splitting (see Sec. \ref{sec:curriculum-training})). And the parameter $K$ is the number of subsets when applying subset extension strategy (see Sec. \ref{sec:curriculum-training}). Therefore, we analyze the performance change on different value settings of $\lambda$, $\theta$ and $K$. 

\begin{table}[tbp]
  \centering
  \caption{Performances under different settings of $\lambda$ on the YOOCHOOSE dataset}
    \begin{tabular}{c|ccccc}
    \toprule
    $\lambda$ & Rec@20 & MRR@20 & Cov@20 & TCov@20 & Tail@20 \\
    \midrule
    1     & 71.44 & 31.92 & 50.85 & 38.96 & 6.73 \\
    2     & 71.32 & 31.91 & 51.71 & 40.02 & 6.83 \\
    5     & 71.27 & 31.89 & 52.31 & 40.77 & 7.01 \\
    10    & 71.17 & 31.85 & 52.70 & 41.29 & 7.12 \\
    50    & 70.68 & 31.63 & 50.32 & 38.38 & 7.40 \\
    100   & 70.12 & 31.28 & 47.98 & 35.40  & 7.03 \\
    200   & 69.73  & 31.10  & 48.00  & 35.45  & 6.80  \\
    500   & 65.00  & 29.07  & 32.61  & 16.43  & 5.41  \\
    \bottomrule
    \end{tabular}
  \label{tab:lambda-YOOCHOOSE}
\end{table}

\begin{table}[tbp]
  \centering
  \caption{Performances under different settings of $\lambda$ on the Last.fm dataset}
    \begin{tabular}{c|ccccc}
    \toprule
    $\lambda$ & Rec@20 & MRR@20 & Cov@20 & TCov@20 & Tail@20\\
    \midrule
    1     & 21.65 & 8.60  & 78.60 & 73.25 & 5.74 \\
    2     & 21.59 & 8.57  & 79.22 & 74.03 & 6.23 \\
    5     & 21.50 & 8.53  & 78.92 & 73.65 & 6.50 \\
    10    & 21.53 & 8.54  & 79.48 & 74.36 & 6.79 \\
    50    & 21.65 & 8.58  & 82.87 & 78.60 & 8.28 \\
    100   & 21.89 & 8.64  & 78.36 & 72.98 & 9.92 \\
    200   & 21.88 & 8.57  & 71.49 & 64.43 & 10.59 \\
    500   & 20.67 & 7.88  & 54.34 & 43.00 & 9.31 \\
    \bottomrule
    \end{tabular}
  \label{tab:lambda-Last.fm}
\end{table}

\subsubsection{The Parameter $\lambda$}\label{sec:lambda}
In this section, we analyze the performance change under different settings of the hyperparameter $\lambda$. We tune $\lambda$ from $\{1, 2, 5, 10, 50, 100,200,500\}$, and the results are shown in Tables \ref{tab:lambda-YOOCHOOSE} and \ref{tab:lambda-Last.fm}.

With the increase in $\lambda$, the accuracy of the recommendation remains stable at the beginning. On the YOOCHOOSE dataset, Rec@20 changes from $71.44\%$ to $71.17\%$ and MRR@20 decreases from $31.92\%$ to $31.85\%$, when $\lambda$ is tuned from 1 to 10. The tendency is similar on the Last.fm dataset. For the long-tail-based metrics, the performances increase with the change in $\lambda$ and decreases when $\lambda$ is set to a large value (e.g., $\lambda=500$). For example, on the YOOCHOOSE dataset, our CSBR(S) model achieves $6.73\%$, $7.4\%$ and $5.41\%$ of Tail@20 when $\lambda =$ 1, 50, and 500, respectively. The larger $\lambda$ can place more emphasis on the calibration loss function (see Eq. \ref{eq:lambda}) in the calibration model. Therefore, Cov@20, TCov@20 and Tail@20 all increase at the beginning. However, an extremely large $\lambda$ causes a negative impact on both accuracy-based and long-tail-based performance. As shown in Tables \ref{tab:lambda-YOOCHOOSE} and \ref{tab:lambda-Last.fm}, the performances of Cov@20, TCov@20 and Tail@20 all decrease when $\lambda=500$. It is normal that overemphasis on calibration can lead to a decrease in prediction accuracy, but the performance of long-tail metrics also declines. One of the possible reasons is that the long-tail performance is also affected by the prediction accuracy. If a model cannot accurately predict the next behavior, it still suffers from popularity bias. Head items are easy to predict because of adequate sequential information, while prediction on tail items distinguishes the abilities of recommendation models. The comparisons among SR-GNN, NARM and GRU4REC in Tables \ref{tab:performance-y4} and \ref{tab:performance-Last.fm} also demonstrate this opinion, where SR-GNN outperforms NARM and GRU4REC in terms of both accuracy-based metrics and long-tail-based metrics. 

\begin{figure}[tbp]
    \centering
    \includegraphics[width=9cm]{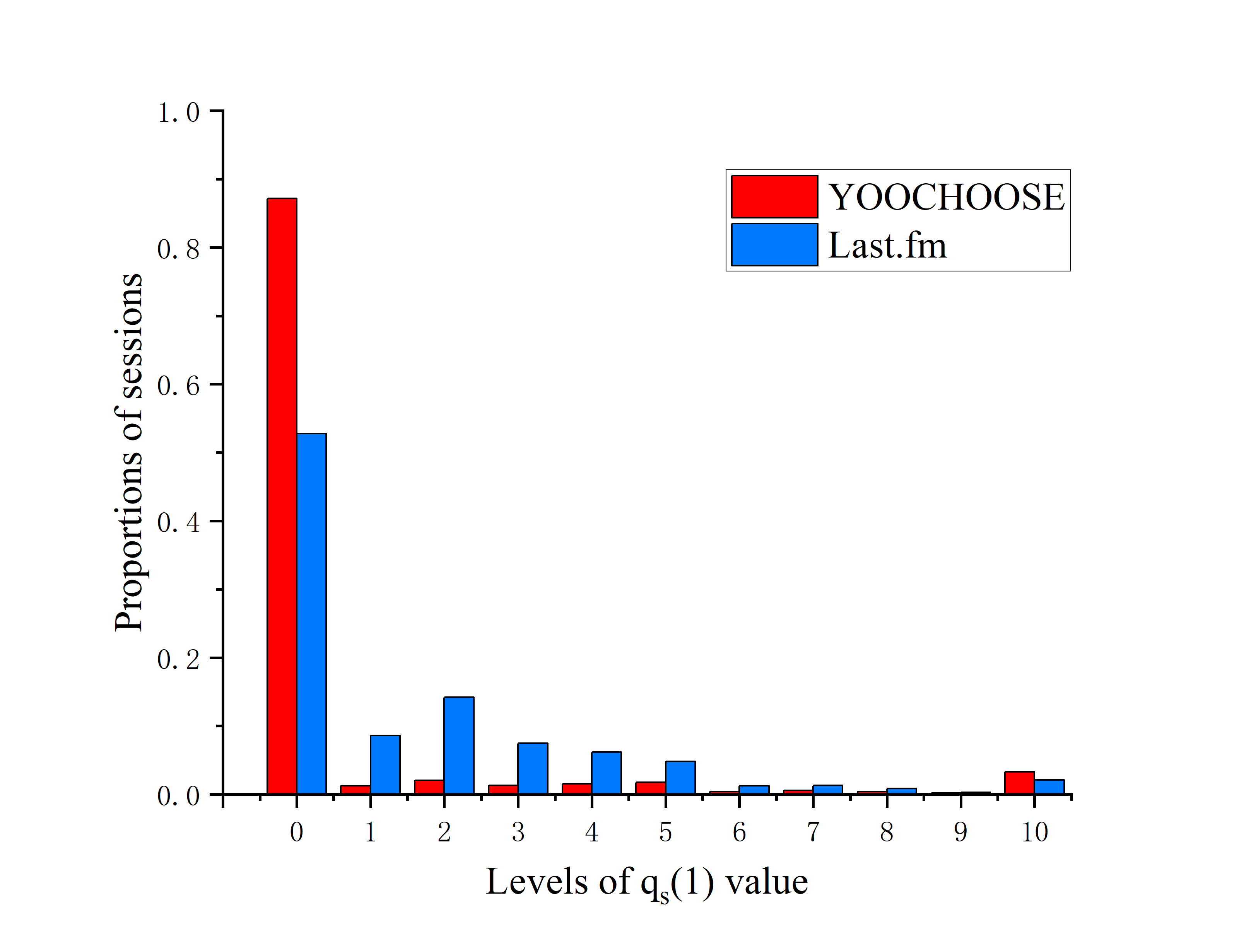}
    \caption{Distributions of sessions according to the $q_s(1)$ levels.}
    \label{fig:session-count}
\end{figure}

\subsubsection{The Parameter $\theta$}\label{sec:theta} 
\begin{figure}[tbp]
\centering
\subfigure[Performances of the accuracy-based Metrics on the YOOCHOOSE dataset]{
\includegraphics[width=5cm]{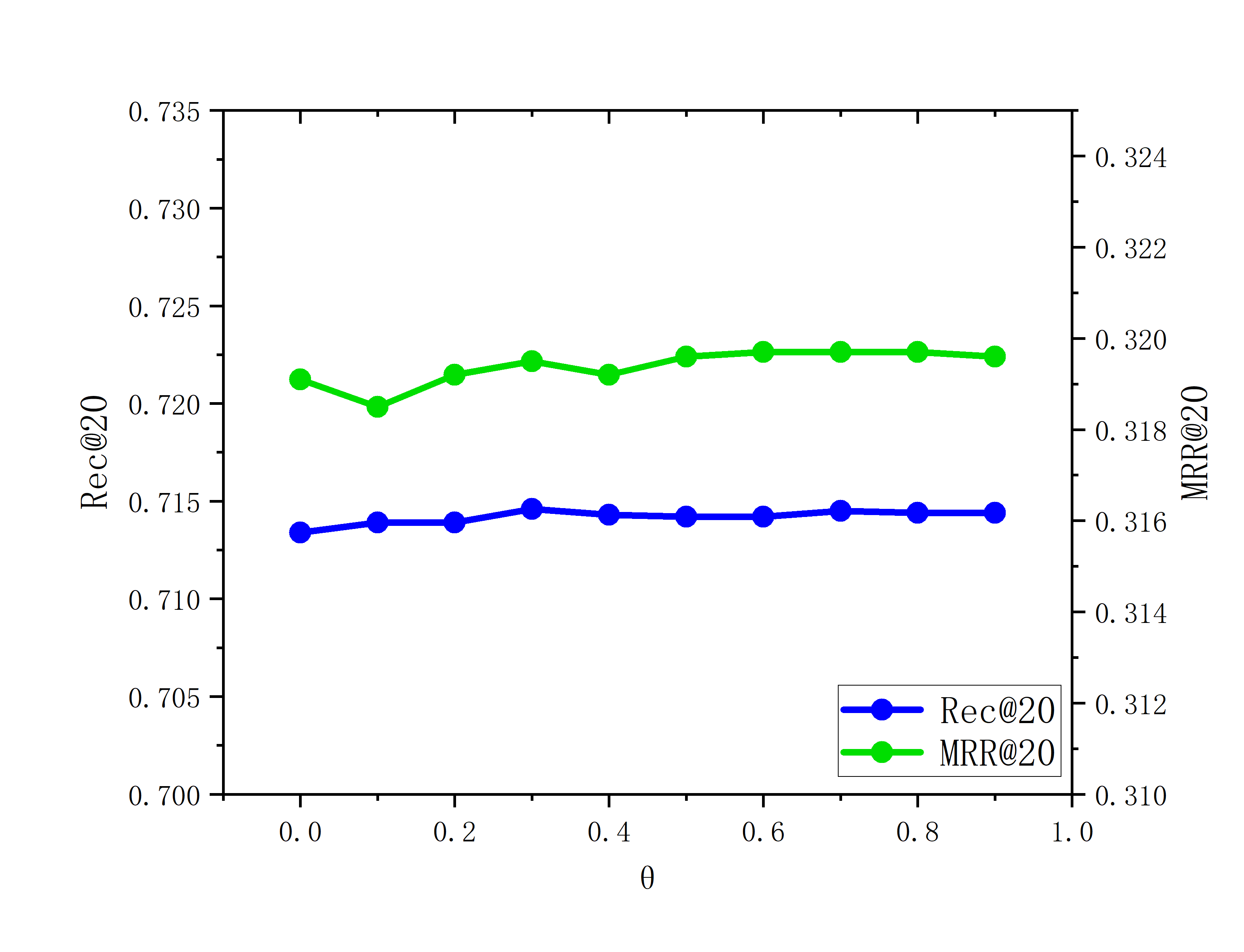}
}
\quad
\subfigure[Performances of the accuracy-based Metrics on the Last.fm dataset]{
\includegraphics[width=5cm]{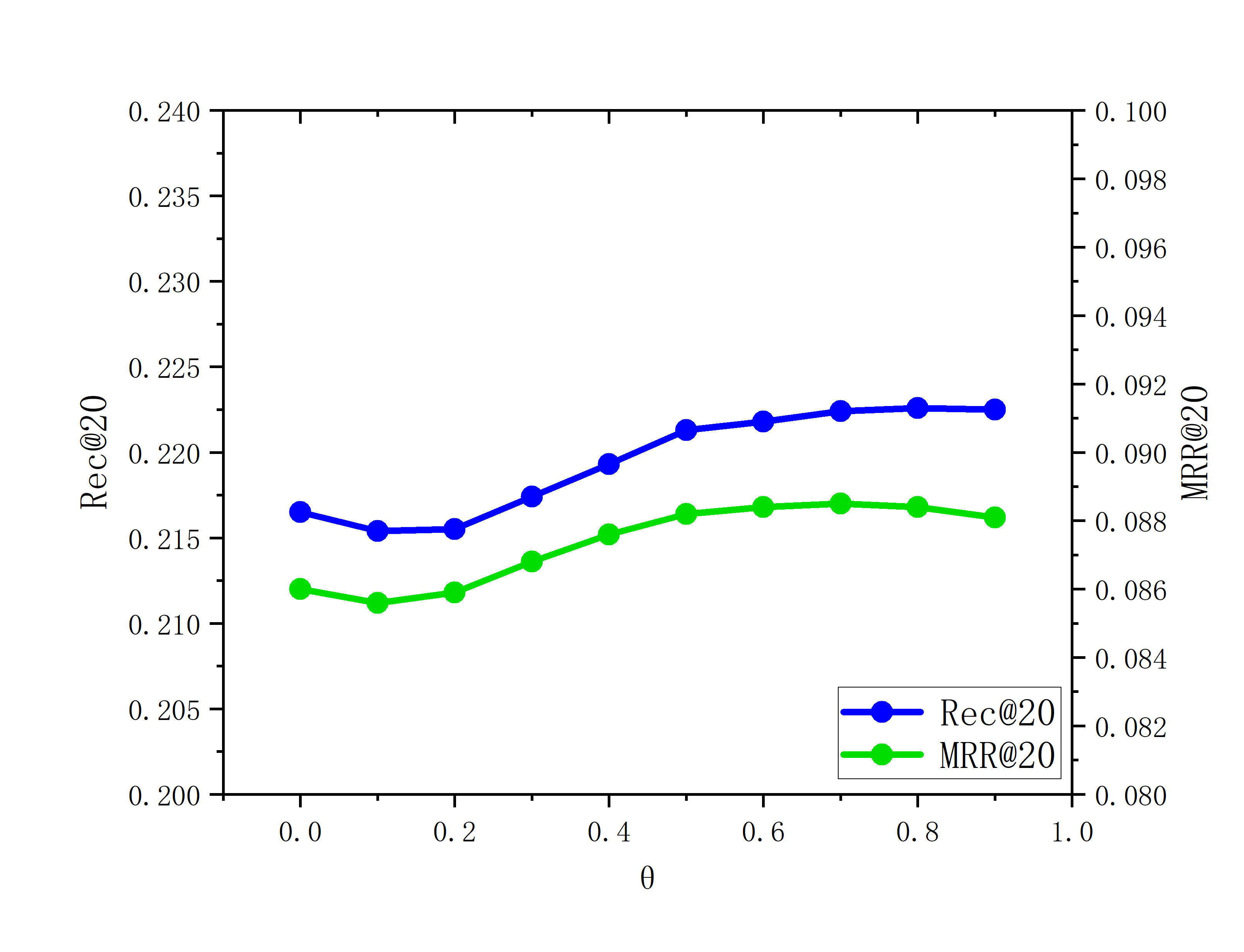}
}
\quad
\subfigure[Performances of the long-tail based Metrics on the YOOCHOOSE dataset]{
\includegraphics[width=5cm]{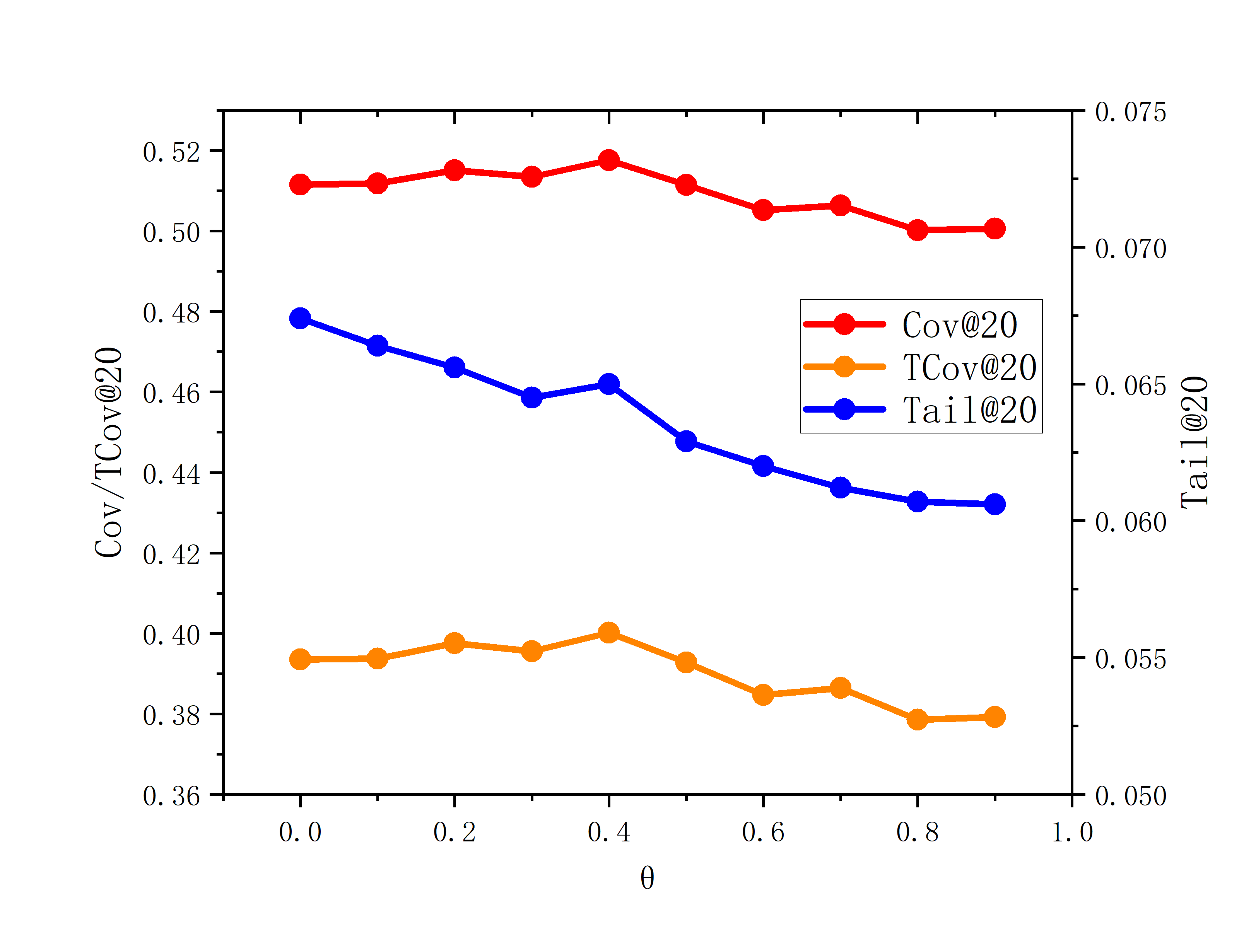}
}
\quad
\subfigure[Performances of the long-tail based Metrics on the Last.fm dataset]{
\includegraphics[width=5cm]{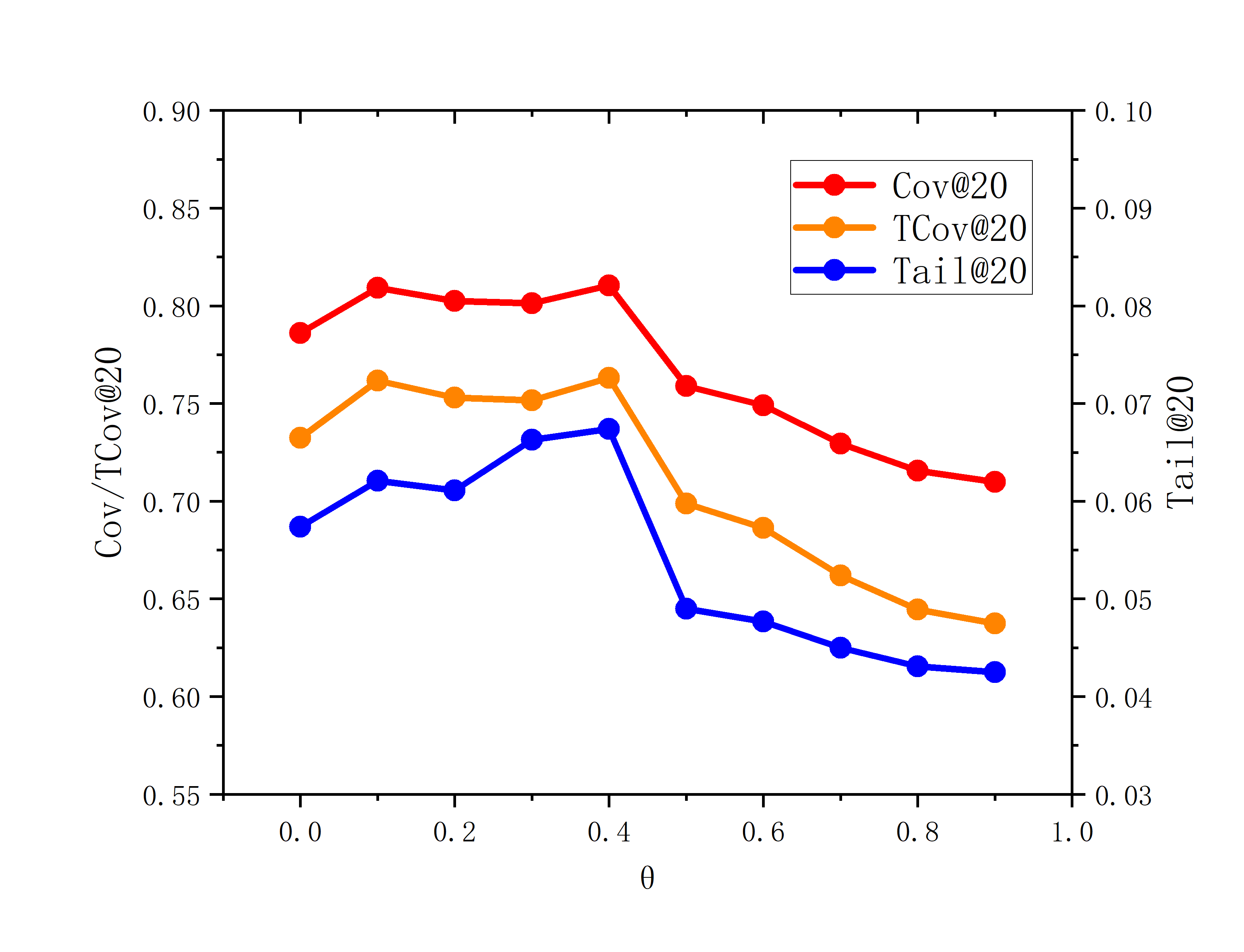}
}
\caption{Performance comparisons when $\theta$ changes.}
\label{fig:theta}
\end{figure}

We also analyze how the performance changes when $\theta$ becomes larger. We make $\theta$ change from 0 to 1, and performances on five metrics are displayed in Fig. \ref{fig:theta}. As shown in Fig. \ref{fig:theta} (a) and Fig. \ref{fig:theta} (b), the recommendation accuracy does not change much, especially on the YOOCHOOSE dataset. This is because of the fixed sequential patterns on the YOOCHOOSE dataset. For long-tail-based metrics, it becomes different. As shown in Fig. \ref{fig:theta} (c) and Fig. \ref{fig:theta} (d), the performances in terms of long-tail-based metrics increase at the beginning of the tuning process on two datasets in general. If we choose a relatively higher value for $\theta$, the long-tail performance decreases dramatically. A possible reason that can explain this phenomenon is the session distribution on the training set. We calculate the proportions of sessions with different levels of the $q_s(1)$ value on the training set. Sessions are divided according to the K-fold extension strategy (see Sec. \ref{sec:curriculum-training}) and K equals to 10. The result is shown in Fig. \ref{fig:session-count}. When $\theta$ becomes much higher, there are fewer sessions for our CSBR model to train. Although the sessions on $S_{te}(Tail)$ can be calibrated to a certain degree, other sessions are sacrificed and the performances decrease. In addition, a high value of $\theta$ makes the subset $S_{te}(Head)$ close to the entire training set, leading to a similar performance of the backbone model. 

On the Last.fm dataset, performances of the long-tail recommendation increase at the beginning, which is different from the YOOCHOOSE dataset. We explain this by the imbalance of session distribution. As shown in Fig. \ref{fig:session-count}, sessions with low $q_s(1)$ occupy the majority on the Last.fm dataset, while the YOOCHOOSE dataset follows a relatively uniform distribution. We think that when the parameter $\theta$ becomes larger, the calibration module can make a better alignment, especially for sessions that contain many tail items because a large number of sessions whose $q_s(1)$ value is lower than $\theta$ are removed. In addition, for these removed sessions, the improvement of our model compared to the backbone model is not as large as the remaining sessions, as shown in Fig. \ref{fig:qlevel}. Therefore, the performance increases at the beginning of the change of parameter $\theta$.

\subsubsection{The Parameter $K$}\label{sec:K} 
Next, we show the performances of our K-fold extension strategy. We change $K$ from $\{1, 2, 4, 8\}$ to observe the performances, where $K=1$ represents the original CSBR(S) model. The results are listed in Table \ref{tab:performance-k}. In general, the accuracy of recommendation declines with the increment of $K$. For example, it changes from $71.44\%$ to $70.52\%$ when $K$ is changed from 1 to 8 on the YOOCHOOSE dataset. In contrast, the ability to recommend tail items is enhanced by increasing $K$. When $K=$ 2 and 4, performances of Cov@20, TCov@20 and Tail@20 are improved compared to the original CSBR(S) model on the two datasets. However, when $K=8$, the model achieves the worst performance in terms of both accuracy and long-tail recommendation. The K-fold extension narrows the range of the target for each separate model and enhances the ability of the calibration module, leading to the improvement of long-tail recommendation. In exchange, the recommendation accuracy is sacrificed. Note that the value of $K$ is not as large as it should be. This is because the distribution labels in each subset become homogeneous, which results in the data imbalance problem. In this situation, the calibration module cannot learn the relationship between the session representation and the ratio of tail items in the recommendation list. In addition, a larger $K$ results in fewer training sessions for each separate session encoders. Therefore, these session encoders fail to learn sufficient information about item transitions, leading to a decrease in the recommendation accuracy.

\begin{table}[tbp]
  \centering
  \caption{Performances of the K-fold extension strategy.}
    \begin{tabular}{c|cccccc}
    \toprule
    Dataset &  K   & Rec@20 & MRR@20 & Cov@20 & TCov@20 & Tail@20 \\
    \midrule
    \multirow{2}[8]{*}{YOOCHOOSE} & 1     & 71.44  & 31.92  & 50.85 & 38.96  & 6.73 \\
       & 2     & 71.13 & 31.84 & 54.06 & 42.93 & 6.82 \\
       & 4     & 70.84 & 31.73 & 55.12 & 44.22 & 6.66 \\
         & 8     & 70.52 & 31.53 & 53.95 & 42.75 & 6.54 \\
    \midrule
    \multicolumn{1}{c|}{\multirow{2}[8]{*}{Last.fm}} & 1     & 21.65 & 8.60   & 78.60 & 73.25 & 5.74 \\
        & 2     & 21.24 & 8.45  & 81.01 & 76.26 & 5.75 \\
         & 4     & 20.51 & 8.19  & 85.47 & 81.84 & 6.92 \\
         & 8     & 20.49     & 8.13     & 77.62     & 72.04     & 4.60 \\
    \bottomrule
    \end{tabular}
  \label{tab:performance-k}
\end{table}

\subsection{RQ4: Ablation Study}

\begin{table}[tbp]
  \centering
  \caption{The performance comparisons of the ablation study on the YOOCHOOSE dataset. }
    \begin{tabular}{c|ccccc}
    \toprule
          & \multicolumn{1}{l}{Rec@20} & \multicolumn{1}{l}{MRR@20} & \multicolumn{1}{l}{Cov@20} & \multicolumn{1}{l}{TailCov@20} & \multicolumn{1}{l}{Tail@20} \\
    \midrule
    CSBR(S) & 71.44 & 31.92 & 50.85 & 38.96 & 6.73 \\
    SR-GNN & 71.45 & 31.85 & 42.35 & 28.29 & 5.66 \\
    w/o Calib & 71.34 & 31.93 & 50.51 & 38.51 & 6.57 \\
    w/o CE & 61.58 & 27.96 & 29.28 & 12.67 & 1.94 \\
    w/o CT & 71.36 & 31.62 & 41.89 & 27.68  & 5.71 \\
    w/o WL & 71.37 & 31.90 & 52.05 & 40.43 & 6.90 \\
    \bottomrule
    \end{tabular}
  \label{tab:ablation-YOOCHOOSE}
\end{table}

\begin{table}[tbp]
  \centering
  \caption{The performance comparisons of the ablation study on the Last.fm dataset. }
    \begin{tabular}{c|ccccc}
    \toprule
          & \multicolumn{1}{l}{Rec@20} & \multicolumn{1}{l}{MRR@20} & \multicolumn{1}{l}{Cov@20} & \multicolumn{1}{l}{TailCov@20} & \multicolumn{1}{l}{Tail@20} \\
    \midrule
    CSBR(S) & 21.65 & 8.60 & 78.60 & 73.25 & 5.74 \\
    SR-GNN & 22.44 & 8.88 & 67.44 & 59.31 & 4.11 \\
    w/o Calib & 21.68 & 8.62 & 75.54 & 69.44 & 5.08 \\
    w/o CE & 11.50 & 4.64 & 29.54 & 12.24 & 6.11 \\
    w/o CT & 22.03 & 8.76 & 67.14 & 58.92 & 4.56 \\
    w/o WL & 21.27 & 8.40 & 70.60 & 63.26 & 5.25 \\
    \bottomrule
    \end{tabular}
  \label{tab:ablation-Last.fm}
\end{table}

In this section, we answer question RQ4 about the contribution of the calibration module and the curriculum training mechanism of our CSBR model. Therefore, we conduct an ablation study of the CSBR(S) model by comparing it with the backbone SR-GNN model and its variants:
\begin{itemize}
    \item w/o Calib: removing the alignment of our model ($\lambda = 0$ in Eq. \ref{eq:lambda}).
    \item w/o CE: removing the cross entropy loss function ($L_{CE}$ in Eq. \ref{eq:lambda}) in our model.
    \item w/o CT: removing the curriculum training strategy (see Sec. \ref{sec:curriculum-training}) in our model, where the calibration module is computed on all sessions.
    \item w/o WL: removing the additional weight (Eq. \ref{eq:weight}) of the loss function for the calibration module.
\end{itemize}
The performance comparisons are shown in Table \ref{tab:ablation-YOOCHOOSE} and Table \ref{tab:ablation-Last.fm}, and we set $\lambda = 1$ and $\theta = 0$. 

The curriculum training contributes more than other components except for the cross entropy in the calibration module in terms of the long-tail-based metrics. Removing the curriculum training strategy results in a similar performance compared to the original backbone model. On the YOOCHOOSE dataset, the Cov@20 of the SR-GNN model is $42.35\%$, while it is $41.89\%$ by removing the curriculum training (i.e., w/o CT) of our CSBR(S) model. On the Last.fm dataset, the performances of Cov@20 are $67.14\%$ and $67.44\%$ for the SR-GNN model and w/o CT of our CSBR(S) model, respectively. For the TCov@20 and Tail@20 metrics, the performance comparisons are similar. This is because most sessions do not have tail items, as shown in Fig. \ref{fig:session-count}. The values of the calibration loss function for these sessions are equal to zero because of the weighting schema based on the ratio of tail items in the session (see Sec. \ref{eq:weight}). If these sessions are included in the training stage, the calibration module cannot learn sufficient information, leading to a similar performance compared to the backbone model.

Under the curriculum training setting, the cross entropy and calibration loss function both contribute to improving long-tail recommendation. On the one hand, the cross entropy loss function ensures the recommendation accuracy when we persuade long-tail recommendations. If we remove the cross entropy loss function, performances decrease dramatically, even if the session encoder is initialized by pre-trained parameters. For example, on the YOOCHOOSE dataset, Recall@20 decreases from $71.44\%$ to $61.58\%$ by removing the cross entropy function, and Cov@20 decreases from $50.85\%$ to $29.28\%$.

On the other hand, removing calibration is a special set of hyperparameter $\lambda = 0$. When $\lambda = 0$, the performance of long-tail recommendation slightly decreases on the YOOHCOOSE dataset. On the Last.fm dataset, the performances of long-tail-based metrics decrease (e.g., Tail@20 decreases from $5.74\%$ to $5.08\%$). A possible reason is the fixed sequential patterns on the YOOCHOOSE dataset (see Sec. \ref{sec:overall}), which leads to high accuracy and low coverage. A lower value of $\lambda$ does not exert sufficient influence on the training process, but enlarging $\lambda$ can improve the performance of long-tail recommendation (e.g., when $\lambda=10$ in Table \ref{tab:lambda-YOOCHOOSE}). On the Last.fm dataset, sequential patterns are not as fixed as on the YOOCHOOSE dataset, leading to a decline in performance by removing the calibration module. Note that removing the calibration module is a special set of $\lambda$. If we consider more about the long-tail recommendation, we can enlarge the value of $\lambda$, which is shown in Sec. \ref{sec:lambda}.

Finally, the weighting schema for the calibration module contributes differently to the two datasets. On the YOOCHOOSE dataset, removing the weight leads to improved performances, while the performances decline on the Last.fm dataset. We explain this according to the imbalance of session distribution in terms of the $q_s(1)$ value (the ratio of long-tail items in the session). As shown in Fig. \ref{fig:session-count}, the YOOCHOOSE dataset follows a relatively uniform distribution, while the distribution on the Last.fm dataset is imbalanced to a certain degree. We add weights to the loss functions in Eq. \ref{eq:weight} to help our model mitigate the effect of the imbalanced distribution of sessions. On the Last.fm dataset, sessions with a high ratio of tail items are emphasized by this schema, resulting in improved long-tail recommendation performances. However, on the YOOCHOOSE dataset, overemphasis on these sessions under a relatively uniform distribution affects the ability of our calibration and alignment module. Therefore, the performances increase when removing the weights on the YOOCHOOSE dataset.

In conclusion, all of these modules contribute to improving our CSBR(S) model. Specifically, the curriculum training strategy helps our model avoid the negative impact caused by the sessions that do not have tail items. Additionally, the calibration module with the cross entropy loss function ensures the coverage of tail items, and the calibration loss functions improve the ratio of tail items recommended to users. Finally, the weighting schema for the calibration loss function contributes differently according to the session distributions.

\section{Discussion}\label{sec:discussion}
In summary, experiments on benchmark datasets show that our CSBR model achieves competitive prediction accuracy with state-of-the-art SBR models, and outperforms these models in terms of long-tail-based metrics. In detail, the calibration module provides recommendation lists that contain more tail items compared to baselines, and the curriculum training strategy helps the calibration module avoid the imbalance problem. 

Although our work achieves improvement on long-tail session-based recommendation, there are some limitations to our work. 
\begin{itemize}
    \item Our work optimizes the recommendation list according to the number of tail items in historical clicks in the session. For sessions that only contain head items, it is contradictory when considering the long-tail recommendation and calibration. The long-tail recommendation requires models to provide more items for users, because users may feel bored after receiving too many head items. However, a calibrated recommendation list should not contain tail items because no tail items have occurred in the session. Because of the contradictory situation, we do not apply our CSBR model to these sessions.
    \item Additionally, for sessions containing a large number of tail items (e.g., $q_s(1) \textgreater 0.7$), our model may not offer a corresponding number of tail items (see Sec. \ref{sec:theta}) because of the constraint of accurate prediction. Although our CSBR model outperforms baselines in terms of long-tail-based metrics and calibration, the performance is still not satisfactory compared to the ratio of tail items in the ongoing session. 
    \item Finally, we proposed a curriculum training strategy in our CSBR model. Specifically, it includes a threshold-based data organization method and a pre-training and fine-tuning paradigm. Despite the effectiveness of our model, the threshold-based data organization is relatively simple. The data splitting strategy also increases the size of the model (e.g., there are $K+1$ separate backbone models when applying K-fold extensions). Moreover, knowledge transfer among each session encoders is not deeply investigated in our model.
\end{itemize}

\section{Conclusion and Future Work}\label{sec:conclusion}
In this paper, we mitigated the popularity bias in session-based recommendation, where items with low popularity were rarely recommended. We handled the long-tail recommendation from the user's perspective and data organization which were not considered in previous studies.
From the user's perspective, we proposed a calibration module to align the ratio of tail items between the recommendation list and the ongoing session. To ensure the effectiveness of the calibration module, we proposed a curriculum training strategy to organize sessions by threshold-based settings for data organization. We evaluated the performances of our model on two benchmark datasets. The experimental results showed that our model achieved competitive prediction accuracy with state-of-the-art models. For long-tail recommendation, our CSBR model enhanced the ability to recommend tail items compared to either backbone models or state-of-the-art baselines. Experiments on sessions that contain tail items demonstrated that our model can provide more calibrated recommendation lists than backbone models, which was the source of performance improvement. We also analyzed the performance change by tuning the hyperparameters. Finally, the ablation studies showed that both the calibration module and curriculum training strategy contribute to the improvement.

For future work, we are committed to addressing some of the shortcomings of our model. We first plan to conduct user studies about recommending tail items for sessions that only contain head items, because it is contradictory between long-tail and calibrated recommendations for these sessions. In this scenario, we aim to provide recommendation lists with long-tail items supported by theory and data.
In addition, for sessions containing a large number of tail items, our model may not offer a corresponding number of tail items. Disentangling users' intentions from the ongoing session is a possible method for improving the calibration and long-tail recommendation for these sessions.
Finally, it is worth investigating the application of the advanced curriculum learning framework. We intend to utilize a dynamic scoring mechanism on training samples rather than setting a threshold to enhance the performance and reduce the size of the model. In addition, applying transfer learning to transfer information among session encoders is also a future direction.

\bmhead{Statements and Declarations}
This work is funded by National Natural Science Foundation of China (under project No. 61907016) and Science and Technology Commission of Shanghai Municipality, China (under project No. 19511120200 and No. 21511100302) for sponsoring the research work.
All authors certify that they have no affiliations with or involvement in any organization or entity with any financial interest or non-financial interest in the subject matter or materials discussed in this manuscript.


\end{document}